\documentclass{nature}

\usepackage{graphicx}
\makeatletter
\let\saved@includegraphics\includegraphics
\AtBeginDocument{\let\includegraphics\saved@includegraphics}
\renewenvironment*{figure}{\@float{figure}}{\end@float}
\makeatother

\makeatletter
\def\@citep@n#1{%
 \begingroup
  \let\@safe@activesfalse\@empty
  \@nocite{#1}% ignores spaces, writes to .aux file, returns #1 in \@no@sparg
  \@tempcntb\m@ne    % \@tempcntb tracks highest number
  \let\@celt\delimiter % an unexpandable, but identifiable, token
  \def\@cite@list{}% % empty list to start
  \let\@citea\@empty % no punctuation preceding first
  \@for \@citeb:=\@no@sparg\do{\@make@cite@list}% make a sorted list of numbers
  \@tempcnta\m@ne    % no previous number
  \mathchardef\@cite@incr\z@ % no previous sequence
  \let\@h@ld\@empty  % nothing held from list yet
  \let\@celt\@compress@cite \@cite@list % output number list with compression
  \@h@ld % output anything held over
 \endgroup
 \@restore@auxhandle
 }
\DeclareRobustCommand{\citep}{%
  \@ifnextchar[{\@tempswatrue\@citey}{\@tempswafalse\@citey[]}}
\def\@citey[#1]#2{\@citep{\@citep@n{#2}}{#1}}
\def\@citep#1#2{\leavevmode \citep@adjust
  \citeleft{ref.#1\if@tempswa\@safe@activesfalse\citemid{#2}\fi
  \spacefactor\@m % punctuation in note doesn't affect outside
  }\citeright
 \@restore@auxhandle}
\def\citep@adjust{\begingroup%
  \@tempskipa\lastskip \edef\@tempa{\the\@tempskipa}\unskip
  \ifnum\lastpenalty=\z@ \penalty\citeprepenalty \fi
  \ifx\@tempa\@zero@skip \spacefactor1001 \fi % if no space before, set flag
  \ifnum\spacefactor>\@m \ \else \hskip\@tempskipa \fi
  \endgroup}
\makeatother

\usepackage{amsmath, amssymb}
\usepackage{physics, siunitx}
\usepackage{url}

\newcommand{\aap}{\textit{Astron. Astrophys.}}
\newcommand{\aj}{\textit{Astron. J.}}
\newcommand{\apj}{\textit{Astrophys. J.}}
\newcommand{\apjl}{\textit{Astrophys. J. Lett.}}

\newcommand{\araa}{\textit{Annual Rev. Astron. Astrophys.}}
\newcommand{\mnras}{\textit{Mon. Not. R. Astron. Soc.}}

\newcommand{\nat}{\textit{Nature}}
\newcommand{\pasj}{\textit{Publ. Astron. Soc. Japan}}

\title{Destruction of the central black hole gas reservoir through head-on galaxy collisions}

\author{Yohei Miki$^1$, Masao Mori$^2$ \& Toshihiro Kawaguchi$^{3,4}$}

\begin{document}

\maketitle

\begin{affiliations}
 \item Information Technology Center, The University of Tokyo, 5-1-5 Kashiwanoha, Kashiwa, Chiba 277-8589, Japan
 \item Center for Computational Sciences, University of Tsukuba, 1-1-1 Tennodai, Tsukuba, Ibaraki 305-8577, Japan
 \item Department of Economics, Management and Information Science, Onomichi City University, Onomichi, Hiroshima 722-8506, Japan
 \item Visiting Scholar, National Astronomical Observatory of Japan, Mitaka, Tokyo 181-8588, Japan
\end{affiliations}

\begin{abstract}
A massive black hole exists in almost every galaxy. 
They occasionally radiate a vast amount of light by releasing gravitational energy of accreting gas, with a cumulative active period of only a few \SI{e+8}{years}, so-called the duty cycle of the Active Galactic Nuclei. 
Namely, many galaxies today host a starving massive black hole. 
Although galaxy collisions have been thought to enhance nucleus activity\cite{Sanders1988,HernquistMihos1995}, the origin of the duty cycle,  especially the shutdown process, is a still critical issue\cite{Cisternas2011}. 
Here we show that galaxy collisions are also capable of \textit{suppressing} black hole fueling by using an analytic model and three-dimensional hydrodynamic simulations, applying the well-determined parameter sets for the galactic collision in the Andromeda galaxy\cite{Miki2014, Miki2016}. 
Our models demonstrate that a central collision of galaxies can strip the torus-shaped gas surrounding the massive black hole, the putative fueling source. 
The derived condition for switching-off the black hole fueling indicates that a significant fraction of currently bright nuclei can become inactive, reminiscent of fading/dying active nucleus phenomena\cite{Lintott2009,Keel2012,Keel2017,VillarMartin2018} associated with galaxy merging events. 
Galaxy collisions may therefore be responsible both for switching-off and turning-on the nucleus activity, depending on the collision orbit (head-on or far-off-centre). 
\end{abstract}

Recent discoveries of very extended (radius $> \SI{10}{kpc}$), photo-ionized regions around galaxies often show the fading/dying activity of Active Galactic Nuclei (AGN) over $\sim \SI{e+5}{years}$ \citep{Lintott2009, Yoshida2002}. 
These galaxies often show kinematic signatures of outflows and morphological signatures of interacting or merging galaxies or postmerger morphologies (tidal tails). 
The extended structures seem to be the remnants of merging or interacting galaxies, with gas accretion activity onto the massive black holes (MBHs) of the central and/or the colliding satellite galaxies\cite{Keel2012,Kawaguchi2014}. 
These galaxies are in general radio-quiet, undermining the possibility that the large-scale gaseous structure is jet-induced. 
Subsolar metal abundances ($\sim \SI{0.5}{Z_\odot}$) are inferred from the line ratios of the gas in the outskirts of these galaxies\cite{VillarMartin2018, Keel2017}. 
The very extended emission-line regions are most likely tidal debris of galactic interactions, illuminated by the past AGN, which has now faded by several orders of magnitudes. 
The above findings hint at a negative effect on the central activity by galaxy merger. 

The study of galaxy mergers in the local universe has advanced rapidly in recent years. 
In the stellar halo of the Andromeda galaxy (M31), the giant southern stream (GSS) is the most prominent tidal debris feature originating from a past galaxy merger\cite{Ibata2001}. 
So far, gravitational $N$-body simulations of a central (head-on) collision between a satellite dwarf galaxy and M31 reproduced the observed features of the GSS\cite{Fardal2007, MoriRich2008}. 
The systematic search for the infalling orbit and the morphology of the progenitor galaxy constrained the possible orbital parameters and physical properties of the progenitor to a narrow range\cite{Miki2014, Miki2016}. 
The massive dwarf galaxy hits and blankets the galactic centre, and the MBH, originally at the centre of the progenitor, is currently wandering in the halo of M31\cite{Miki2014, Kawaguchi2014}. 
The disc of the dwarf galaxy must have anti-clockwise rotating in the sky to reproduce the observed asymmetric feature along the eastern edge of the GSS\cite{Kirihara2017}. 

\begin{figure}
  \centering
  \includegraphics[width=\columnwidth, pagebox=cropbox, clip]{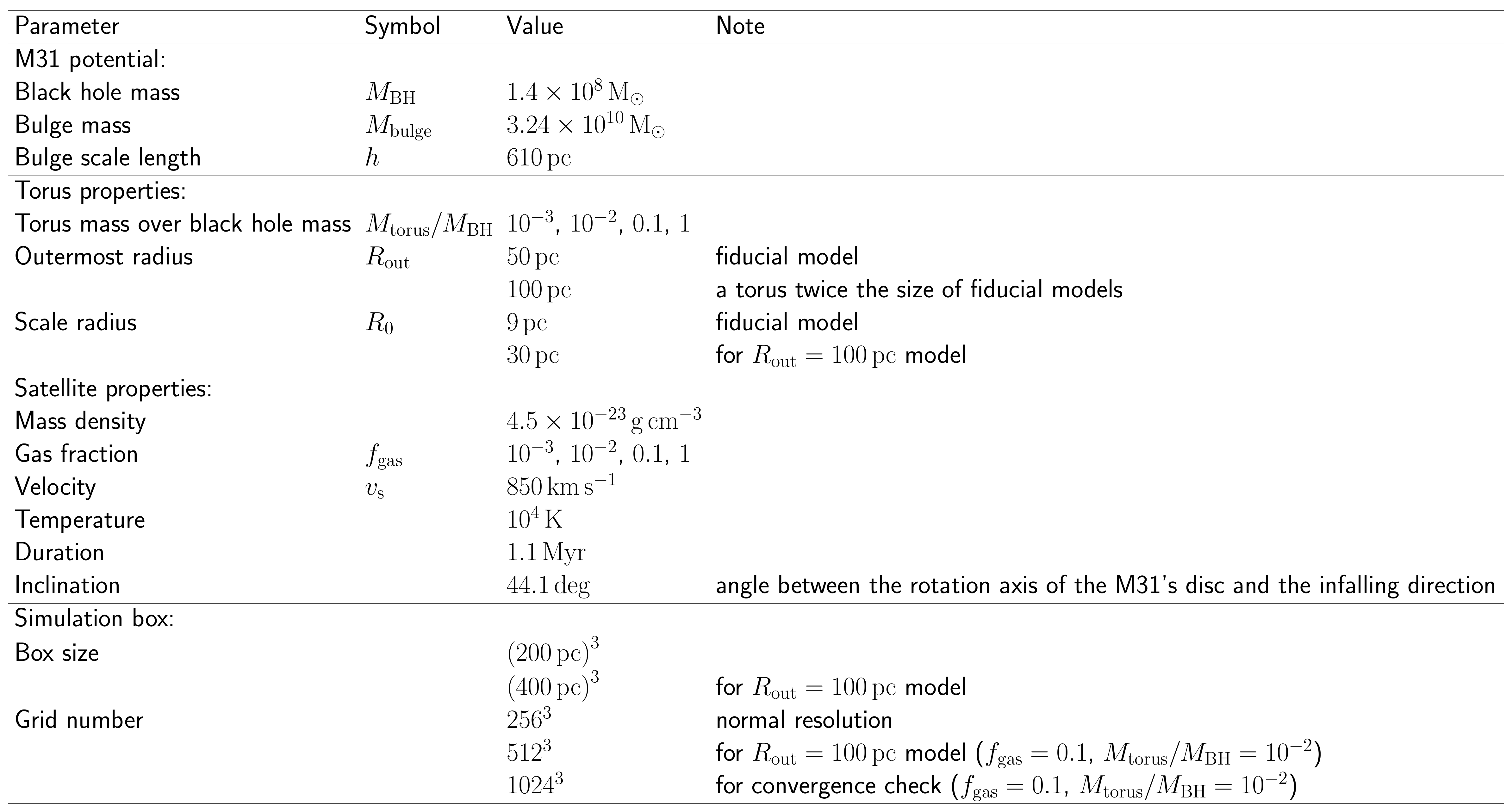}
  \caption{
    \textbf{Simulation parameters.} 
  }
  \label{tab:parameters}
\end{figure}

In this work, we investigate the possibility that central galaxy collisions diminish AGN activity.
Firstly, we focus on M31, which harbours a radiatively quiescent central MBH with a mass $M_\mathrm{BH} = \SI{1.4e+8}{M_\odot}$ \citep{Bender2005} and a quite low X-ray luminosity ($L_{\SIrange[range-phrase=-,range-units=single]{0.3}{7}{keV}} \leq 10^{-10}$ times the Eddington luminosity\cite{Li2009}, a maximum luminosity beyond which radiation pressure will overcome gravity). 
The reason for its extremely low activity, with respect to other inactive galaxies\cite{Ho2009}, is still unknown. 
We use hydrodynamical calculations on \SI{100}{pc} scales (see Methods and Extended Data Figure~\ref{tab:parameters}), assuming that the activity is fuelled by a torus surrounding the central MBH. 
We introduce a critical parameter $\eta \equiv f_\mathrm{gas} \times \pqty{M_\mathrm{torus} / M_\mathrm{BH}}^{-1}$, where $f_\mathrm{gas}$ is the gas fraction of the infalling satellite and $M_\mathrm{torus}$ is the mass of the torus gas. 
Given the sizes of the torus and the satellite, $\eta$ is proportional to the gas column-density ratio of the two bodies. 

\begin{figure*}
    \centering
    \includegraphics[width=\linewidth, pagebox=cropbox, clip]{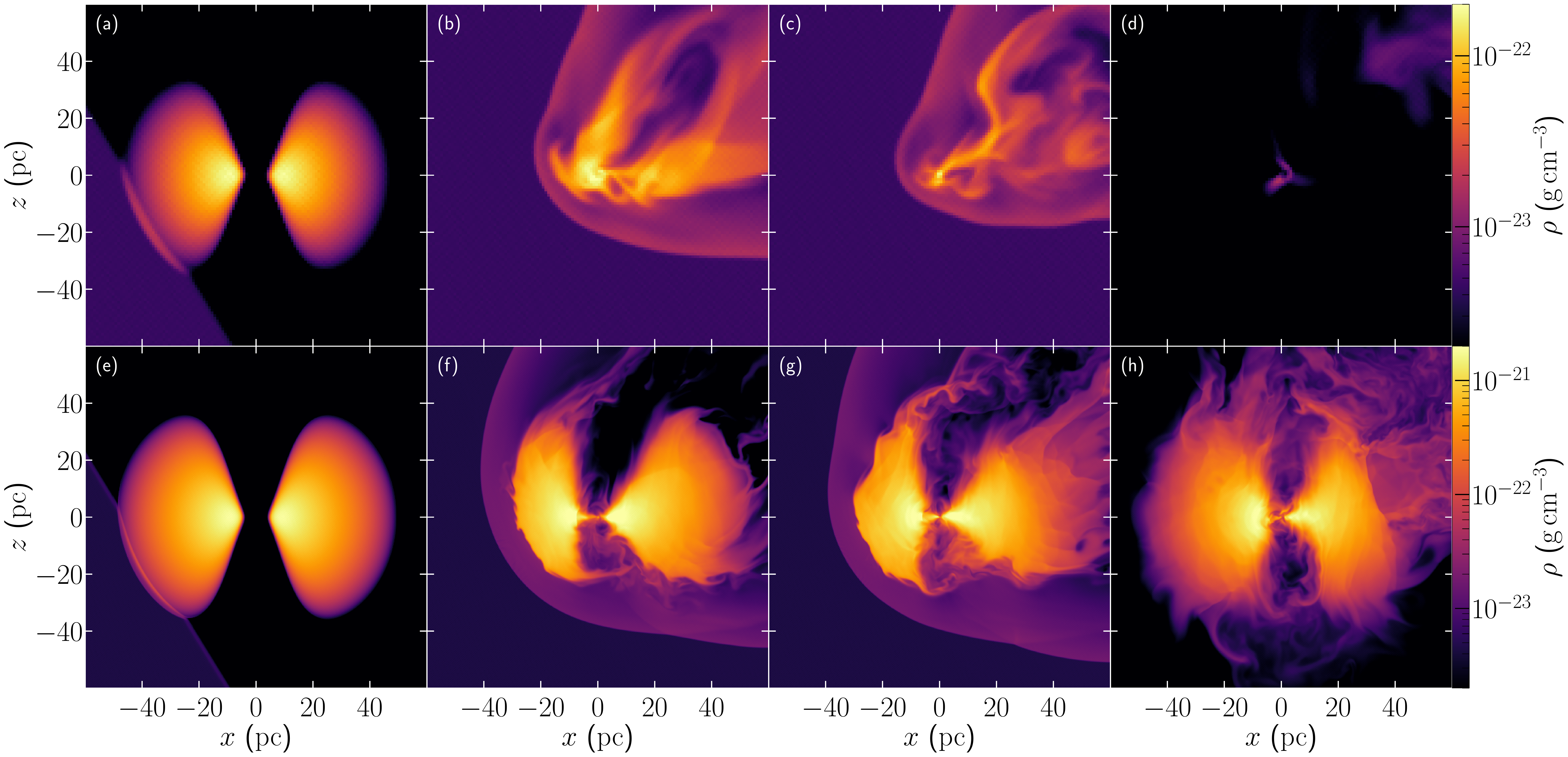}
    \caption{
      \textbf{Time evolution of the torus gas in the meridian plane ($y = 0$) of the torus frame.} 
      Panels a to d and e to h show results for $M_\mathrm{torus} / M_\mathrm{BH} = 10^{-3}$ and $10^{-2}$, respectively, with $f_\mathrm{gas} = 0.1$. 
      From a to d and from e to h, snapshots at \SI{0.02}{Myr}, \SI{0.46}{Myr}, \SI{0.90}{Myr} and \SI{1.36}{Myr} after the gas inflow begins. 
      The colour scale shows the volume-density of the gas. 
      A steady and uniform gas inflow lasting for $\SI{1.1}{Myr}$ representing the infalling satellite hits the torus-shaped gas from the bottom-left corner of the panels, and the temperature of the torus-shaped gas increases to $\sim \SI{e+6}{K}$. 
    }
    \label{fig:time.evolution}
\end{figure*}

Figure~\ref{fig:time.evolution} displays the time evolution of the torus density distribution in the meridian plane for $\eta = 100$ (upper panels) and $\eta = 10$ (lower panels). 
For the $\eta = 100$ model, the propagation of the satellite gas causes intense compression of the torus gas. 
The torus gas is then accelerated by the infalling gas and is entirely pushed out of the gravitational potential well within a short time scale ($< \SI{1}{Myr}$). 
For $\eta = 10$ model, the evolution of the torus gas is completely different from the case of $\eta = 100$. 
The gas is mildly ablated due to Kelvin--Helmholtz instabilities. 
As a result, there are many small irregularities in the surface of the torus gas. 
As the analytic model described in Methods predicts, the torus gas is completely stripped away from the central MBH for $\eta \gtrsim 100$ but survives for $\eta \lesssim 10$. 

\begin{figure*}
  \centering
  \includegraphics[width=.8\columnwidth, pagebox=cropbox, clip]{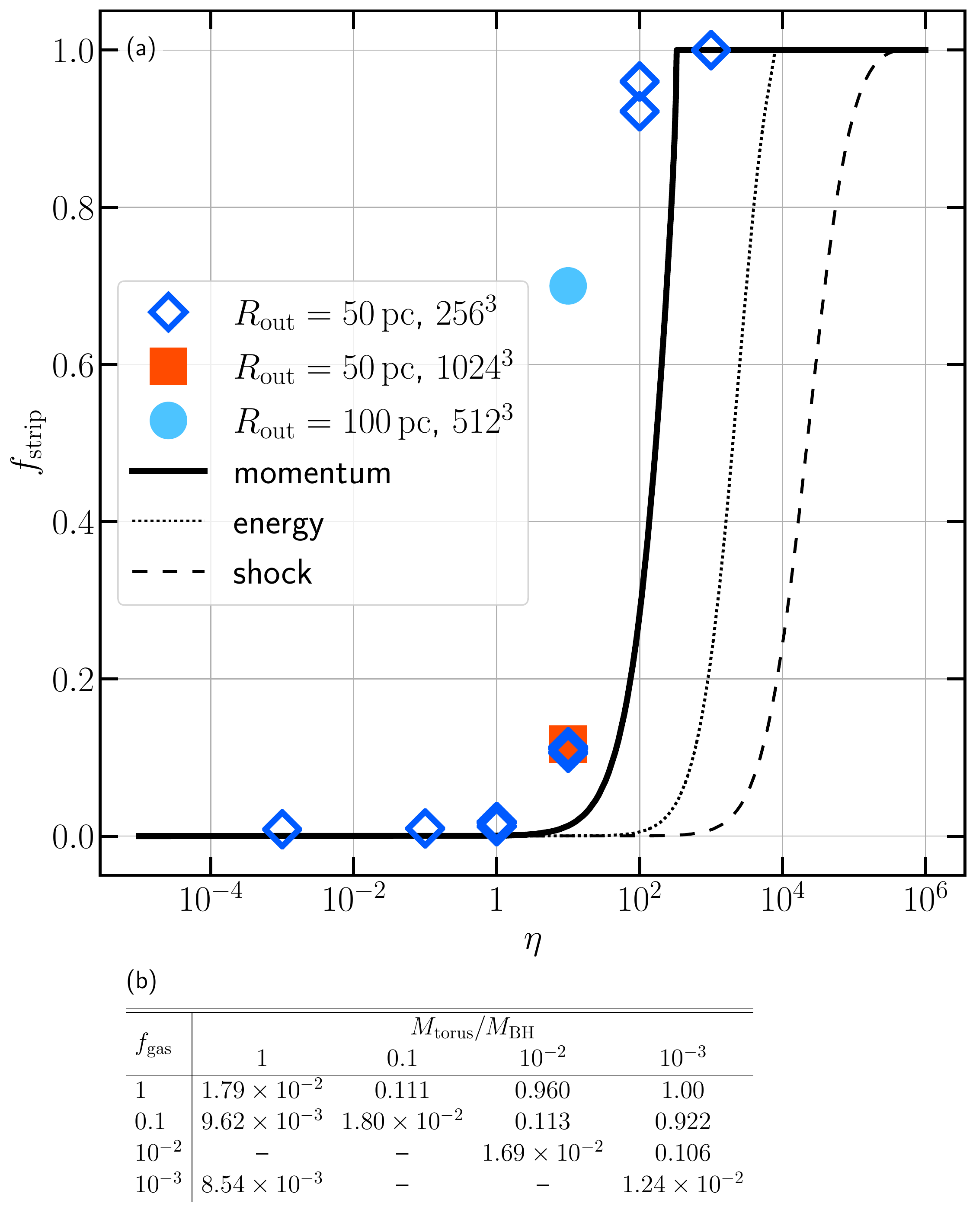}
  \caption{
    \textbf{Mass stripped fraction $f_\mathrm{strip}$.} 
    Panel a shows $f_\mathrm{strip}$ as a function of the column-density ratio parameter $\eta$. 
    The solid, the dotted and the dashed curves represent the analytic estimates (for $R_\mathrm{out} = \SI{50}{pc}$) of the mass stripped fractions based on the momentum transfer (eq.~\ref{eq:momentum.transfer}), the energy (eq.~\ref{eq:torus.energy}) and the velocity (eq.~\ref{eq:torus.velocity}), respectively. 
    The overlaid symbols show results of numerical simulations: the blue diamonds ($R_\mathrm{out} = \SI{50}{pc}$ runs with $256^3$ grid points), the red square (high-resolution run with $1024^3$ grid points: $f_\mathrm{gas} = 0.1$, $M_\mathrm{torus} / M_\mathrm{BH} = 0.01$ and $f_\mathrm{strip} = 0.117$) and the sky-blue circle ($R_\mathrm{out} = \SI{100}{pc}$ run with $512^3$ grid points). 
    Panel b lists $f_{\rm strip}$ as a function of the gas fraction of the infalling satellite, $f_\mathrm{gas}$, and the mass ratio of the torus and the central MBH. 
  }
  \label{fig:ResultsLinear}
\end{figure*}

\begin{figure}
  \centering
  \includegraphics[width=\columnwidth, pagebox=cropbox, clip]{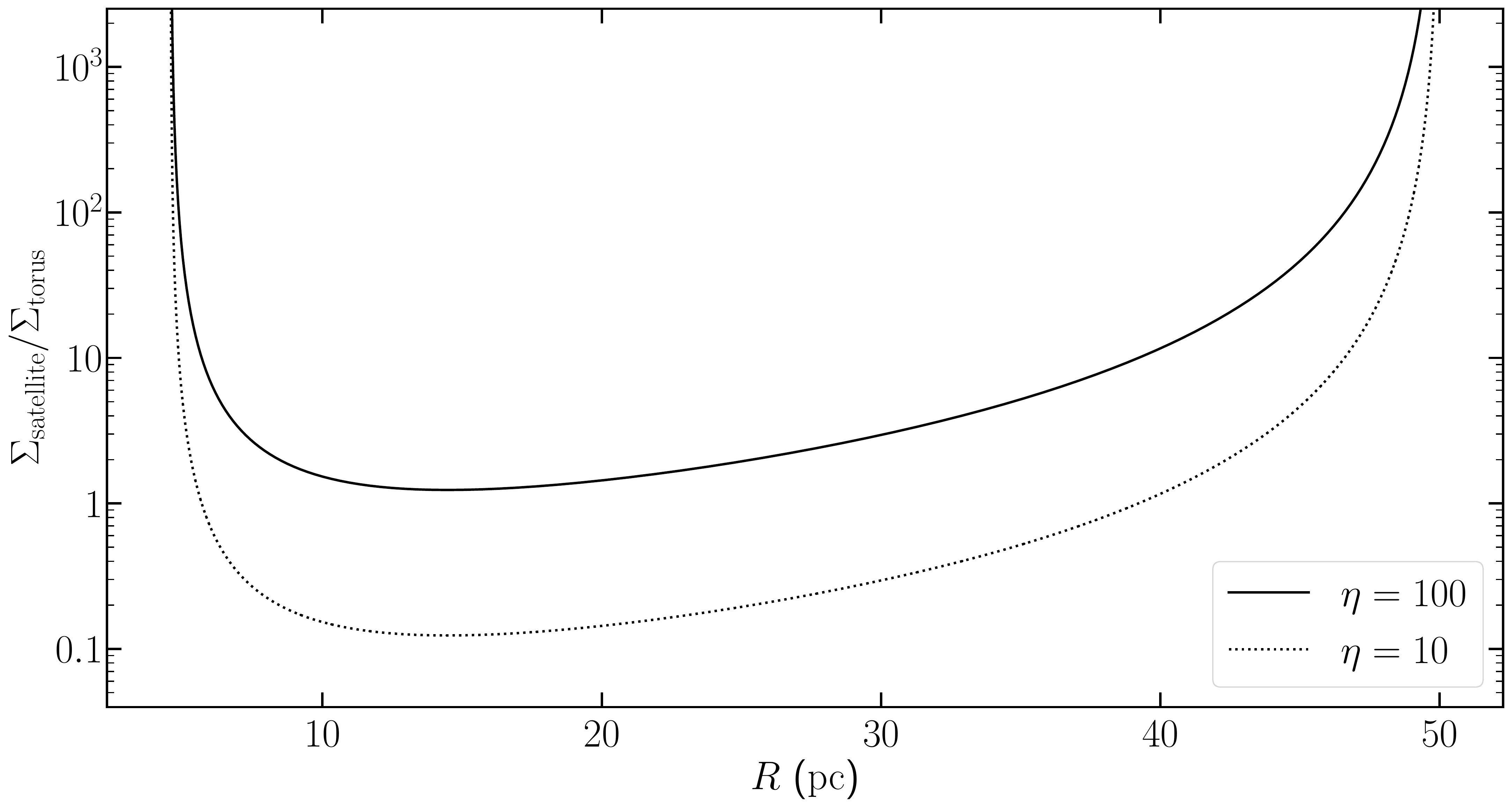}
  \caption{
    \textbf{Radial profile of the ratio of the gas column-densities of the infalling satellite galaxy to that of the central torus, $\Sigma_{\rm satellite} / \Sigma_{\rm torus}$.} 
    The solid and dotted curves show the gas column-densities ratio for $\eta = 100$ and $10$, respectively. 
  }
  \label{fig:ColumnDensityRatio}
\end{figure}

Figure~\ref{fig:ResultsLinear}a compares the numerical results (symbols) with the analytic model predictions (curves) as a function of $\eta$, and Figure~\ref{fig:ResultsLinear}b lists the stripped fraction of the torus gas $f_\mathrm{strip}$ (see Methods). 
The result is almost identical to that in the normal resolution run ($256^3$ grid points), indicating convergence with spatial resolution. 
Figure~\ref{fig:ResultsLinear}a shows how a sudden change of the stripped fraction occurs within the range of $10 \lesssim \eta \lesssim 100$ as predicted by the analytical model (solid curve in Figure~\ref{fig:ResultsLinear}a). 
It implies that a large amount of the torus gas is swept out via momentum transfer when the column-density of the torus gas is lower than that of the satellite galaxy (see also Extended Data Figure~\ref{fig:ColumnDensityRatio}). 
Although both calculations show that the essential parameter driving the gas removal is the gas column-density ratio, $\eta$, numerical simulations tend to strip more gas away from the central MBH than the analytical estimation. 

There is, therefore, a critical value of $\eta$ ($\eta \gtrsim 25$), beyond which the mass fueling onto the central MBH may be inhibited due to a diminished fuel reservoir. 
In contrast, if the column-density of the torus gas is high enough, there is little effect on the structure of the torus. 

The stripped fraction for a torus double the radius, $R_\mathrm{out} = \SI{100}{pc}$, but the same mass (for which we also double the adopt box size and the grid point numbers) is $0.700$ (sky-blue circle in Figure~\ref{fig:ResultsLinear}a). 
The larger torus size results in a smaller column-density $\Sigma_\mathrm{torus}$, hence leading to the higher stripped fraction. 
It confirms that the critical condition to suspend the MBH fueling by central collisions of galaxies is the gas column-density ratio. 
These results show that the condition for gas stripping is as follows: 
\begin{equation}
    M_\mathrm{gas} \gtrsim \SI{3e+7}{M_\odot} \times \pqty{\frac{M_\mathrm{torus}}{\SI{e+5}{M_\odot}}} \times \pqty{\frac{R_\mathrm{out}}{\SI{50}{pc}}}^{-2} \times \pqty{\frac{r_\mathrm{gas}}{\SI{1}{kpc}}}^{2}
    . 
  \label{eq:strip_condition}
\end{equation}
Here, $M_\mathrm{gas}$ and $r_\mathrm{gas}$ are the mass and the radius of the gaseous components in the infalling satellite galaxy. 
An equivalent formulation using the column-density ratio is: the column-density of the infalling gas, $\Sigma_\mathrm{gas}$, must exceed that of the torus. 
The larger torus and the smaller satellite galaxy lead to easier gas stripping and effective suppression of MBH fueling. 

\begin{figure*}
  \centering
  \includegraphics[width=.8\columnwidth, pagebox=cropbox, clip]{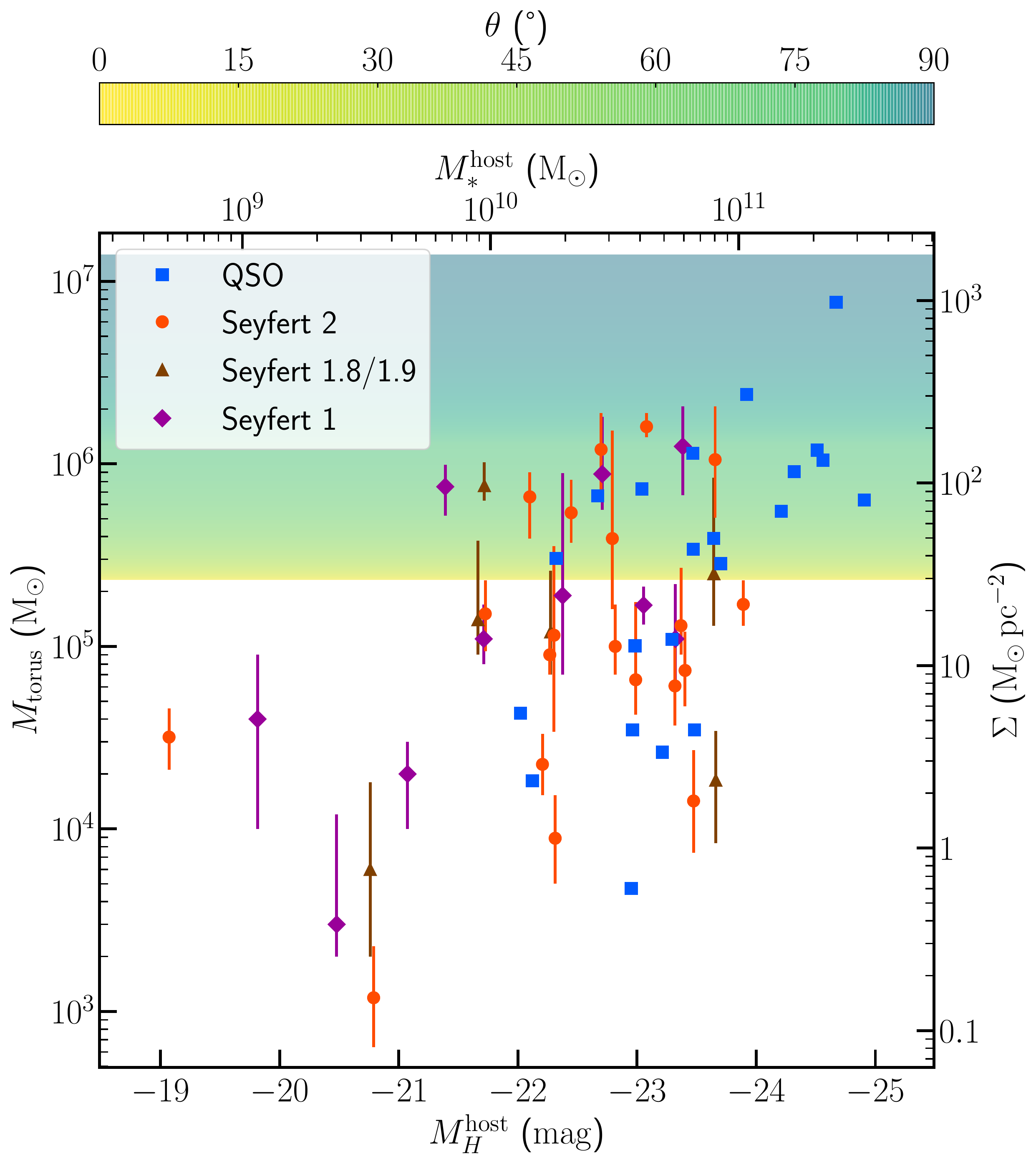}
  \caption{
    \textbf{Relation between the torus mass (with uncertainties indicating the $1 \sigma$ confidence level) and the host galaxy (absolute magnitude in $H$ band, $M_H^\mathrm{host}$, and the corresponding stellar mass $M_*^\mathrm{host}$; see Methods) of AGN.} 
    Blue squares and other symbols correspond to QSOs and Seyfert galaxies (red circles, brown triangles and purple diamonds for type 2, type 1.8/1.9 and type 1 Seyfert galaxies, respectively). 
    The right axis is column-density either for torii or infalling satellites. 
    In converting the gas mass to the column-density of torii, we use the approximation as $M_\mathrm{torus} / \pi R^2_\mathrm{out}$ assuming $R_\mathrm{out} = \SI{50}{pc}$. 
    The shaded zone indicates the stripping criterion (eq.~\ref{eq:strip_condition}) for various orientation angles $\theta$ between torus and satellite rotation axis. 
    See Methods for detailed estimation of $M_\mathrm{torus}$ and the column-density of the infalling gas $\Sigma_\mathrm{gas}$. 
  }
  \label{fig:Mtorus-Mhost}
\end{figure*}

So far, we have focused on a specific parameter setting for the central collision that occurred in M31. 
We expect, however, that central collisions commonly occur in many galaxies, including host galaxies of AGN. 
We now translate the critical condition for AGN shut-off (eq.~\ref{eq:strip_condition}) to a condition involving AGN parameters. 
In Figure~\ref{fig:Mtorus-Mhost}, we superimpose in colour the condition onto a plot of the absolute magnitude of a sample of host galaxies in the $H$ band v.s. the torus mass (see Methods). 
Below the critical zone (yellow to cyan coloured patch), where we find a significant fraction of AGN are located, the AGN activity can cease by a central collision of a satellite galaxy.

Namely, we here propose that galaxy collisions are potentially the essential process to control the fate of AGN activity. 
Galaxy collisions can trigger a starburst and gas flows towards central MBHs, via angular momentum transfer, and trigger an AGN\cite{Sanders1988}. 
On the other hand, luminous merging galaxies, such as ultra-luminous infrared galaxies, do not always harbour AGN\cite{Imanishi2010}. 
Signatures of major mergers in host galaxies of AGN as seen by the Hubble space telescope images show no clear systematic differences from those in inactive galaxies\cite{Cisternas2011}. 
Although evidence of multiple episodes in AGN activity has hitherto been accumulated\cite{Saripalli2003}, 
what controls the AGN activity (e.g., ignition and duration) is still unclear. 
Especially, the shutdown process of AGN activity has not yet concluded, 
while various mechanisms (e.g., accretion towards the central MBH, AGN feedback, and galactic bar resonance, see Methods) were proposed. 

In Figure~\ref{fig:Mtorus-Mhost}, the width of the critical zone reflects the uncertainty of the density distribution of the gas disc in the infalling satellite and the orientation angle $\theta$ between torus and satellite axis of rotation. 
Edge-on infall increases the column-density of the infalling gas and hence enhances the torus stripping compared to the face-on infall. 
Fainter galaxies appear to have torii that are more susceptible to the interaction of the infalling satellite, and therefore the AGN activity would be easily shut-off; reminiscent of AGN preferentially observed in massive galaxies\cite{Yamada2009}. 
Edge-on views of these torii would correspond to heavily obscured (so-called Compton-thick) AGN. 
Tidal compression of the infalling satellite, not considered here, would enlarge the critical zone by increasing the column-density of the infalling gas. 

\begin{figure*}
    \centering
    \includegraphics[width=.8\columnwidth, pagebox=cropbox, clip]{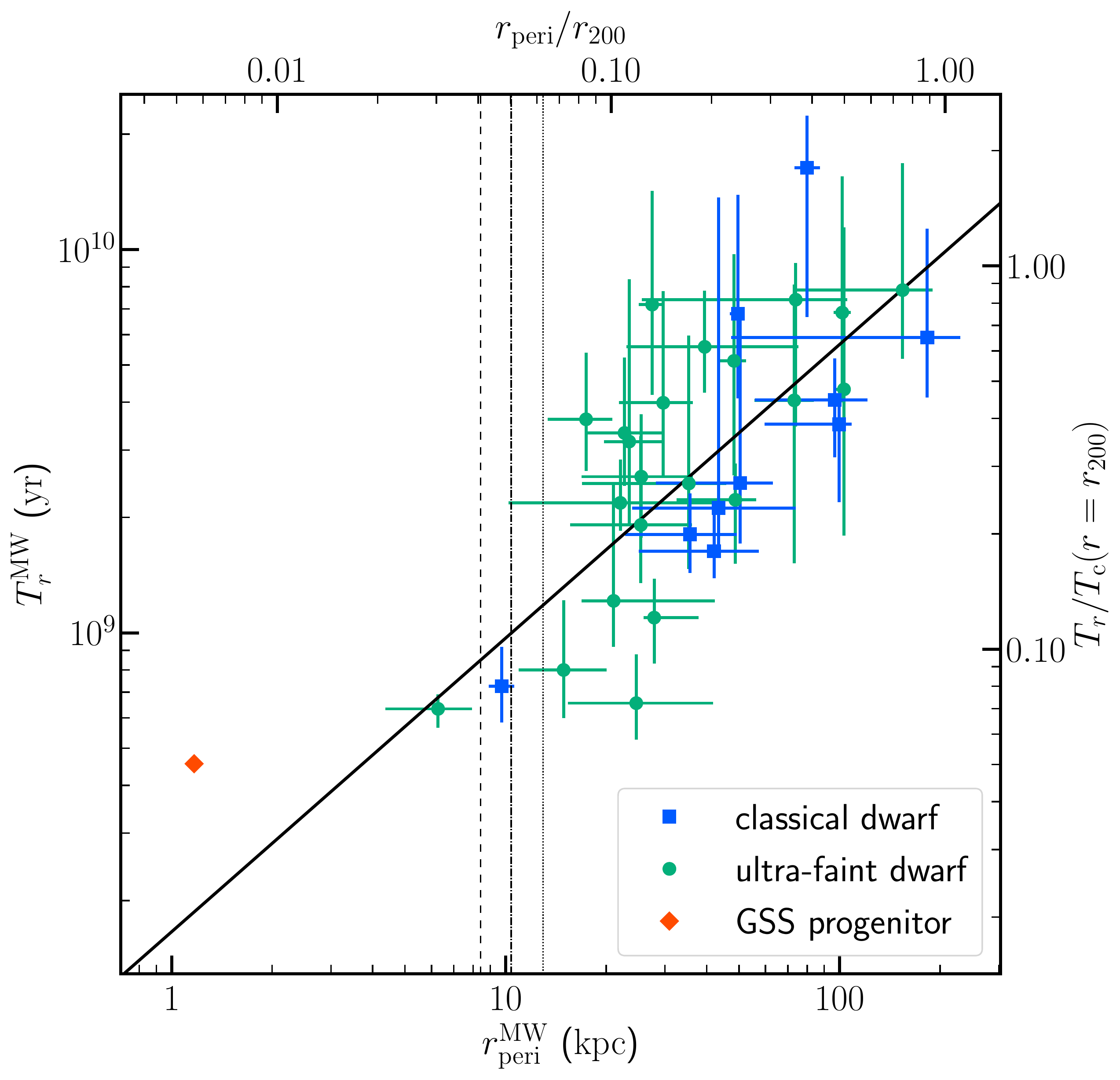}
    \caption{
      \textbf{Distribution of orbital periods, $T_r$, and pericentres, $r_\mathrm{peri}$, for the satellite galaxies in the Milky Way (blue boxes for classical dwarfs and green circles for ultra-faint dwarfs; uncertainties indicate the $1 \sigma$ confidence level).} 
      The solid line represents the best-fit model for the Milky Way satellites: $T_r = 170 \pqty{r_\mathrm{peri} / \SI{1}{kpc}}^{0.77}$~\si{Myr}. 
      The vertical lines at $r_\mathrm{peri} \sim \SI{10}{kpc}$ indicate the tidal-stripping radii for \SIlist{e+7;e+8;e+9}{M_\odot} haloes from left to right (see Methods). 
      For comparison, the progenitor for the GSS around M31 is plotted in red, using the right and top axes. 
      The top and right axes are normalized by the virial radius, $r_{200}$, and the orbital period there: $T_\mathrm{c} = \SIlist[list-units = single, list-pair-separator = {,}]{9.1; 8.6}{Gyr}$ at $r_{200} = \SIlist[list-units = single, list-pair-separator = {,}]{207; 195}{kpc}$ for Milky Way and M31, respectively. 
    }
    \label{fig:GaiaDR2}
\end{figure*}

\begin{figure}
  \centering
  \includegraphics[height=\textheight, pagebox=cropbox, clip]{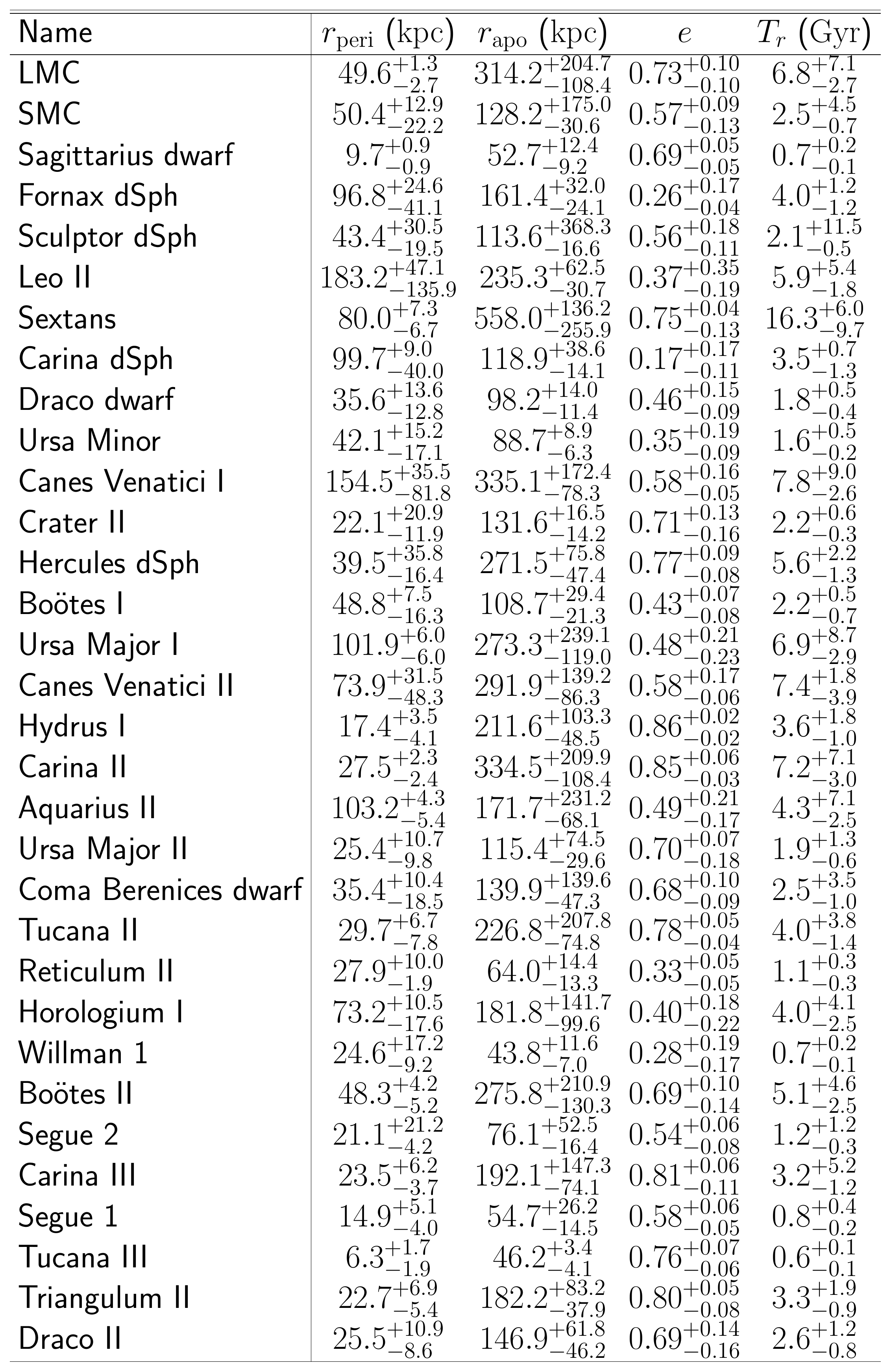}
  \caption{
    \textbf{Orbital parameters of the Milky Way satellites.} 
  }
  \label{tab:MWsat}
\end{figure}

The hierarchical clustering model based on the standard cold dark matter cosmology predicts that the orbital eccentricity distribution of sub-galactic dark matter haloes in a Milky Way-sized dark matter halo is radially biased relative to that of all subhaloes associated with the host halo\cite{Benson2005, Khochfar2006, Barber2014}. 
The radial orbit of the sub-galactic haloes can not only inhibit the mass fueling in the galactic centre but also lead to subhalo depletion, as a consequence of tidal destruction in the proximity of the galaxy centre ($\lesssim \SI{15}{kpc}$) \citep{Garrison-Kimmel2017}. 
Using the most recent proper motion measurements provided by the second data release of the \textit{Gaia} mission (\textit{Gaia} DR2), we integrated orbits of the satellite galaxies surrounding the Milky Way (see Methods and Extended Data Figure~\ref{tab:MWsat}). 
The distributions of orbital periods, $T_r$, and periapsises, $r_\mathrm{peri}$, in Figure~\ref{fig:GaiaDR2} show a linear (in logarithmic space) trend from the upper-right towards lower-left. 
Cosmological simulations indicate that about \numrange[range-phrase=--]{10}{30} subhaloes in a Milky Way-sized halo have been stretched and disrupted by the tidal force of the central galaxy, and the debris has fallen into the galactic centre. 
For instance, \num{10} subhaloes with gas, at $r_{\rm peri} \lesssim \SI{15}{kpc}$ and orbital period $\lesssim \SI{e+9}{years}$, would reduce nuclear MBH fueling in the central galaxy every $\sim \SI{e+8}{years}$ on average, at the expense of the infalling subhalo itself. 
A recent high-resolution cosmological $N$-body simulation implies that thousands of subhaloes penetrated the central region ($\lesssim \SI{10}{kpc}$) in the nine Milky Way-sized haloes\cite{Morinaga2019}. 
The inferred average merger rate over the cosmic age (1 per $\order{\SI{e+8}{yr}}$ per a single host halo) is consistent with our estimation. 
In addition, \textit{Gaia} DR2 recently provided evidence for a central collision event in the Milky Way. 
The Gaia--Enceladus--Sausage is identified as the remnant of a central collision that the Milky Way experienced about $\SI{e+10}{years}$ ago with a massive satellite galaxy\cite{Helmi2018, Belokurov2018}. 
As previously stated, the GSS is the remnant of a recent central collision in M31. 
The GSS progenitor had a smaller pericentre and shorter orbital period compared to the Milky Way satellites known in \textit{Gaia} DR2. 
Besides the GSS, the \SI{10}{kpc} ring, a ring structure observed in M31, may be another signature of a recent central collision that occurred in M31\cite{Gordon2006}. 

An important point to be emphasised is that only central collisions of galaxies open up this new channel for suppressing central MBH fueling. 
Far-off-centre collisions/encounters are thought to occasionally enhance AGN activity triggering the gas flows to the central MBH via angular momentum transfer\cite{HernquistMihos1995, RamonFox2020}. 
Once gas fueling is triggered, however, the nuclear activity could be suppressed by the central collisions, which happen every $\sim \SI{e+8}{years}$, being consistent with the current estimate of the AGN durations. 
If orbital properties of satellite galaxies make a critical impact on the turning on/off of MBH fueling, merging events of galaxies may play an essential role in the AGN duty cycle. 

\begin{figure}
    \centering
    \includegraphics[width=.6\columnwidth, pagebox=cropbox, clip]{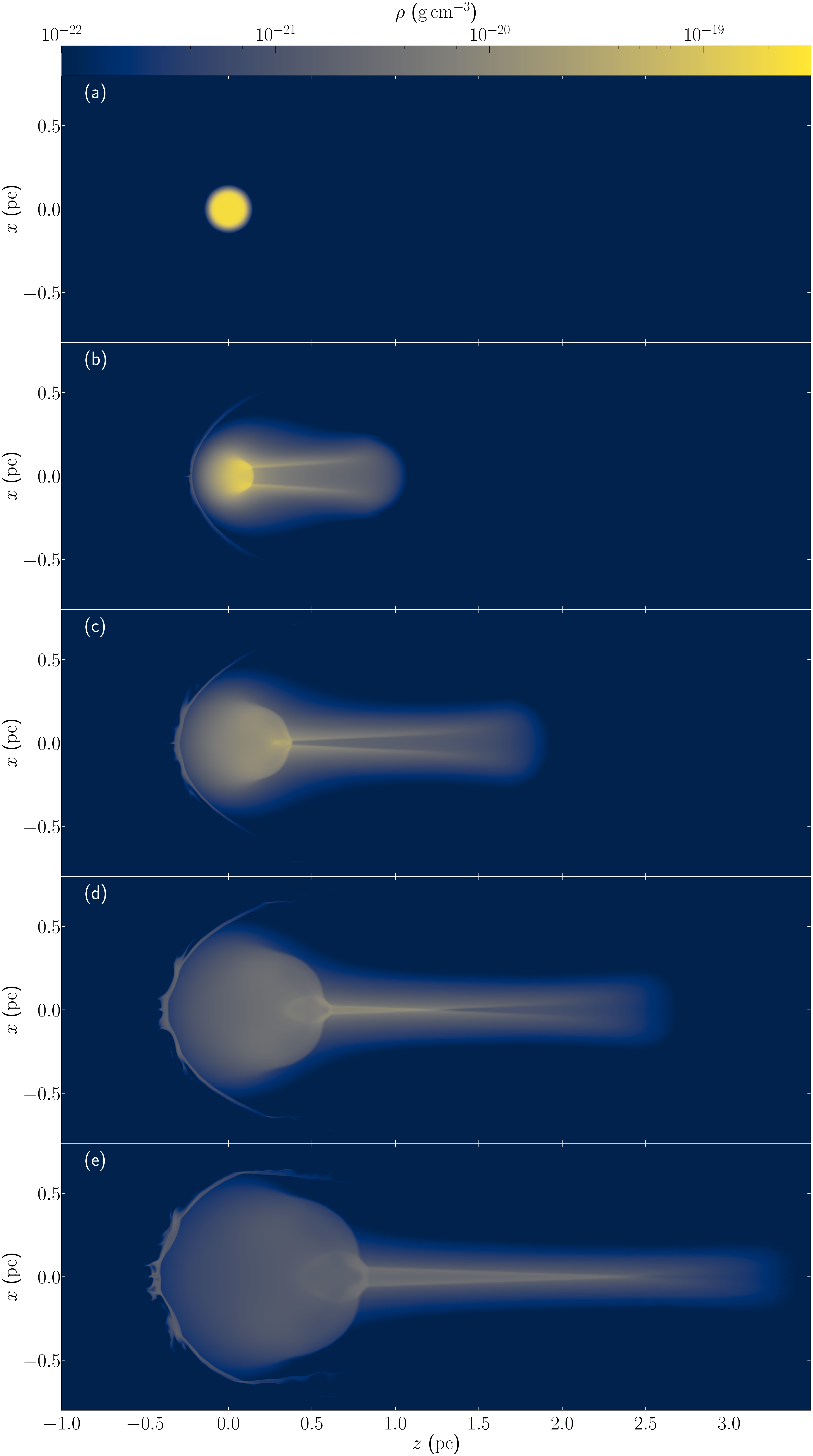}
    \caption{
      \textbf{A series of snapshots of a clump exposed to the fast gas travelling from the left side of the simulation box.} 
      From a to e, each panel shows the volume-density distribution $\rho(z, x)$ in the $y = 0$ plane at $0$, $1$, $2$, $3$, $4$~\si{kyr} after the simulation starts. 
    }
    \label{fig:clump}
\end{figure}

\begin{figure}
    \centering
    \includegraphics[width=\columnwidth, pagebox=cropbox, clip]{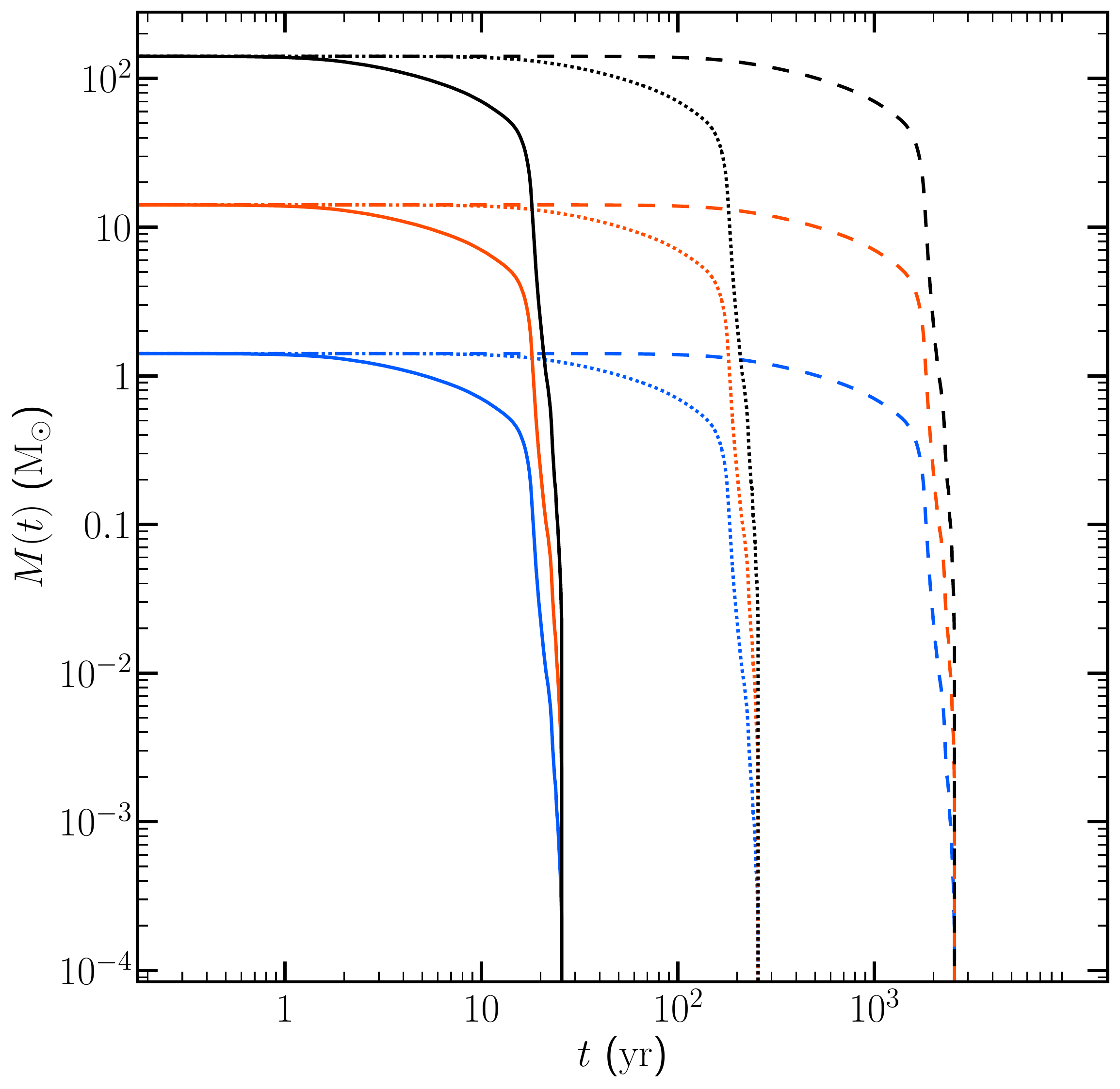}
    \caption{
      \textbf{Time evolution of various clumps exposed to fast infalling gas.} 
      Different colours and linetypes indicate various clump masses $M_\mathrm{clump}$ and clump radii ($r_\mathrm{clump}$): $M_\mathrm{clump} = \SI{1.4}{M_\odot}$ (blue), $\SI{14}{M_\odot}$ (red) and $\SI{140}{M_\odot}$ (black); $r_\mathrm{clump} = \SI{e-3}{pc}$ (solid), $\SI{e-2}{pc}$ (dotted) and $\SI{0.1}{pc}$ (dashed). 
    }
    \label{fig:massloss}
\end{figure}

\begin{figure}
    \centering
    \includegraphics[width=\columnwidth, pagebox=cropbox, clip]{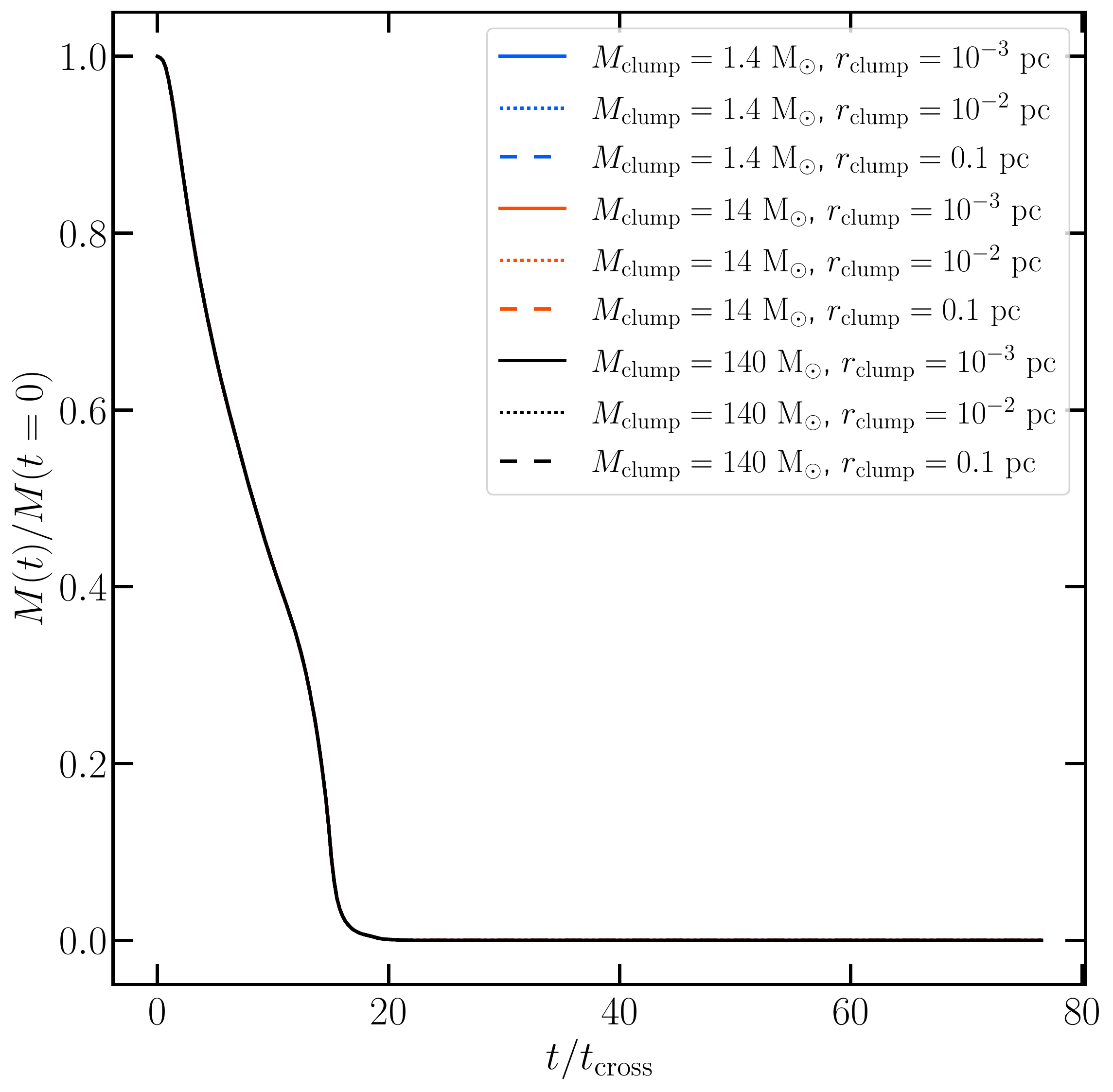}
    \caption{
      \textbf{Time evolution of various clumps examined in Extended Data Figure~\ref{fig:massloss}, presented with the normalised quantities in both axes.} 
      The crossing time of a clump $t_\mathrm{cross}$ is defined as $r_\mathrm{clump}$ divided by the infalling gas flow velocity of $\SI{850}{km.s^{-1}}$: $t_\mathrm{cross} = \SI{1.2}{yr}$, $\SI{12}{yr}$ and $\SI{120}{yr}$ for $r_\mathrm{clump} = \SI{e-3}{pc}$, $\SI{e-2}{pc}$ and $\SI{0.1}{pc}$, respectively. 
      All nine curves are totally overlapping in the normalised axes. 
    }
    \label{fig:massloss-scaled}
\end{figure}

In this \textit{Letter}, we assume a smooth torus, although the torus may in fact be clumpy (see Methods and Extended Data Figures~\ref{fig:clump}, \ref{fig:massloss} and \ref{fig:massloss-scaled}). 
Uncertainties and dependencies on various collision parameters (such as infalling direction, velocity and pericentric distance) are not considered in this work. 
It is also worth investigating how long this suppression mechanism of the AGN activity lasts. 
Future studies utilising detailed simulations with a larger computational box for a longer duration, including gas inflow towards the torus from the galactic disc, self-gravity of the torus gas and radiative cooling processes, will reduce such uncertainties.

%% References

\begin{addendum}
 \item
    We are grateful to Alex Wagner, Masayuki Umemura and Ken Ohsuga for careful reading of the manuscript and for their comments that improved the manuscript. 
    Numerical computations were performed with computational resources provided by Multidisciplinary Cooperative Research Program in Center for Computational Sciences, University of Tsukuba, Oakforest-PACS operated by the Joint Center for Advanced High-Performance Computing (JCAHPC) and resources in the Information Technology Center, The University of Tokyo. 
    This work was supported in part by Grants-in-Aid for Specially Promoted Research by MEXT (16002003) and Grant-in-Aid for Scientific Research (S) by JSPS (20224002), (A) by JSPS (21244013), (C) by JSPS (JP17K05389) and JSPS KAKENHI Grant Numbers JP20K14517, JP20K04022. 

 \item[Author Contributions]
   YM contributed to modeling, code development, numerical simulations, analysis, discussion and manuscript preparation; MM contributed to modeling, connection to recent Galactic observations, discussion and manuscript preparation; TK contributed to modeling, generalization to actual AGN environments, discussion and manuscript preparation. 

 \item[Competing Interests]
   The authors declare that they have no competing financial interests. 

 \item[Correspondence]
   Correspondence and requests for materials should be addressed to YM (email: ymiki@cc.u-tokyo.ac.jp). 
\end{addendum}

%%
%% FIGURES in main text
%%

%% Methods
\begin{methods}

\subsection{Earlier studies focusing on AGN activity.}
Many studies have continuously investigated the shutdown mechanism of AGN activity. 
Accretion towards the central black hole reduces the torus gas, and it leads the depletion of the fueling source. 
However, the lifetime of clumpy torii is estimated to be a few to \SI{10}{Myr} \citep{BeckertDuschl2004}, which is much longer than the duration of the simulations in this work. 
It means that the accretion in the torus does not influence the torus destruction process during the central collision of galaxies, focused in this work. 
AGN feedback (via jet or radiative processes) may destroy the dusty torus. 
Numerical simulations, including strong AGN feedback, demonstrated the regulation of the gas fueling to the central region under the assumption of isotropic feedback\cite{DiMatteo2005}. 
High-resolution, radiation hydrodynamic simulations showed that the AGN feedback leads bipolar outflows having little influence in the gas fueling\cite{GaborBournaud2014}. 
Other possible channels for the AGN shutdown are the termination of the gas fueling process from the galactic disc via the galactic bar and/or the Lindblad resonance. 
These processes will need dynamical timescale at a galactic scale (i.e., kpc scale) to vary their efficiency; therefore, they seem to be longer than the lifetime of AGN events (e.g., the bar destruction takes a few Gyr \citep{Athanassoula2005}). 

Merger-triggered enhancement of the AGN activity has been investigated.  Numerical simulations showed that the galaxy collisions triggered mass fueling to the central region and suggested AGN activity enhancement\cite{DiMatteo2005, HopkinsQuataert2010}. 
Recent high-resolution simulations modelling multi-phase interstellar medium with incorporating realistic physical processes resolve the mass fueling down to sub-parsec scales\cite{Angles-Alcazar2020}.

\subsection{Setup for M31 collision simulations.}
\label{sec:setup}
We describe the model assumptions and parameters for the gas around M31's centre and the infalling satellite. 
We focus on the central region of the galaxy, which is dominated by the MBH and the bulge: the gravitational potential of the MBH and a bulge (with a mass $M_\mathrm{bulge} = \SI{3.24e+10}{M_\odot}$ and the scale length $h = \SI{610}{pc}$ \citep{Geehan2006}) are modelled as a fixed point mass and Hernquist potential: 
\begin{align}
    \Phi(r) = -\frac{G M_\mathrm{BH}}{r'} - \frac{G M_\mathrm{bulge}}{r' + h},
\end{align}
where $r'$ is calculated as $\sqrt{\pqty{\SI{e-8}{pc}}^2 + x^2 + y^2 + z^2}$ to remove the divergence due to division by zero. 
This treatment is sufficiently accurate for the central \SIrange{10}{100}{pc} of M31. 

Before the collision, the torus is modelled as an equilibrium, axisymmetric polytrope gas with a heat capacity ratio $\gamma = 5 / 3$ under the spherical gravitational field $\Phi(r)$ as a function of the three-dimensional distance $r$ from the centre. 
The gas density $\rho$, the effective potential $\Phi_\mathrm{eff}$ and the specific angular momentum $l$ in cylindrical coordinates are as follows\cite{TomisakaIkeuchi1988, Okada1989, FukueSanbuichi1993, Bulbul2010}: 
\begin{align}
    \rho(R, z) &= \rho_0 \pqty{\frac{C - \Phi_\mathrm{eff}(R, z)}{C - \Phi_\mathrm{eff}(R, z = 0)}}^{1 / (\gamma - 1)}
    \label{eq:torus.density}
    ,\\
    \Phi_\mathrm{eff}(R, z) &= \Phi(r) + \frac{l^2(R)}{2 R^2}
    \quad
    \textrm{and}\quad
    l(R) = \sqrt{R^3 \pdv{\Phi(r)}{R}}
    . 
\end{align}
We further assume that $l(R)$ is constant over radii\cite{Okada1989, Bulbul2010}, 
\begin{equation}
    l(R) = \sqrt{R_0^3 \eval{\pdv{\Phi(R, z = 0)}{R}}_{R = R_0}}. 
\end{equation}
The constant $C$ in equation~(\ref{eq:torus.density}) is determined by requiring $C = \Phi_\mathrm{eff}(R = R_\mathrm{out}, z = 0)$ at the outermost radius $R_\mathrm{out}$ of the torus gas. 
Here, we assume $R_\mathrm{out} = \SI{50}{pc}$ \citep{Hoenig2006} as a fiducial value. 
The maximum value of the aspect ratio, the scale height of the torus $H(R)$ over $R$, should be about unity\cite{KrolikBegelman1988}. 
Setting the scale radius $R_0$ to be \SI{9}{pc} and $R_\mathrm{out} = \SI{50}{pc}$ meets this requirement. 
The mass of the torus gas $M_\mathrm{torus}$ is in the range of $10^{-3} \leq M_\mathrm{torus} / M_\mathrm{ BH} \leq 1$ \citep{Mor2009,Schartmann2005}. 

The total mass, the scale radius, the velocity and the periapsis of the infalling satellite are \SI{3e+9}{M_\odot}, \SI{1}{kpc}, \SI{850}{km.s^{-1}} and \SI{1}{kpc} \citep{Miki2014, Miki2016}, respectively. 
Since the progenitor of the GSS interacts with the central region of M31 over a crossing time of \SI{1.1}{Myr}, the gas of the infalling satellite is modelled as gas inflow lasting for \SI{1.1}{Myr}. 
The central mass-density of the infalling satellite is estimated to be \SI{4.5e-23}{g.cm^{-3}} \citep{Miki2016}. 
We assume the gas density of the infalling satellite is $4.5\times10^{-23} f_\mathrm{gas}$~\si{g.cm^{-3}}, where $f_\mathrm{gas}$ is the gas fraction in the range of $10^{-3} \leq f_\mathrm{gas} < 1$ \citep{Mateo1998, Conselice2003, Thuan2016}. 
The gas temperature of the infalling satellite is assumed to be \SI{e4}{K}.

\subsection{One-dimensional analytic estimations.}
\label{sec:1D}
We estimate the stripped fraction of the torus gas due to the hydrodynamic interaction with the infalling gas in the satellite galaxy analytically. 
For simplicity, we consider the case in which the satellite moves along the rotation axis of the torus gas ($z$-axis) and solve the one-dimensional problem along the $z$-axis. 
The initial volume-density of the torus gas is set to be $\overline{\rho}_\mathrm{torus}(R) = \int_0^{H(R)} \dd{z} \rho(R, z) / H(R)$. 
The initial vertical velocity of the torus is assumed to be zero. 

Solving the one-dimensional Riemann problem yields the velocity $v_\mathrm{t}$, pressure $p_\mathrm{t}$ and density $\rho_\mathrm{t}$ of the torus gas just after shock passage. 
Based on previous studies of the GSS, we determine whether the gas is stripped or still bound at \SI{10}{kpc} ($r_\mathrm{max}$) from the centre, at which the free-fall time is about \SI{100}{Myr}. 

We set three independent and necessary conditions for gas stripping to estimate the stripped fraction of the torus gas. 
The first condition is that the velocity of the torus gas after the collision exceeds the escape velocity: 
\begin{equation}
    \frac{1}{2} v_\mathrm{t}^2 + \Phi(R, z = 0) \geq \Phi(r_\mathrm{max}). 
    \label{eq:torus.velocity}
\end{equation}
The second condition is that the sum of the kinetic and thermal energy of the torus gas exceeds the potential difference: 
\begin{equation}
    \frac{1}{2} v_\mathrm{t}^2 + \frac{1}{\gamma - 1} \frac{p_\mathrm{t}(R)}{\rho_\mathrm{t}(R)} + \Phi(R, z = 0) \geq \Phi(r_\mathrm{max}).
    \label{eq:torus.energy}
\end{equation}
In this case, the stripped torus gas will re-accrete within the cooling timescale. 
The third condition is the momentum transfer: 
\begin{equation}
    \frac{1}{2} V_\mathrm{both}^2 + \Phi(R, z = 0) \geq \Phi(r_\mathrm{max}),
    \label{eq:momentum.transfer}
\end{equation}
where 
\begin{equation}
    V_\mathrm{both} \equiv \frac{\Sigma_\mathrm{torus} v_\mathrm{t} + \Sigma_\mathrm{satellite} v_\mathrm{s}}{\Sigma_\mathrm{torus} + \Sigma_\mathrm{satellite}}.
\end{equation}
The stripped fraction of the torus gas, $f_\mathrm{strip}$, for each condition is defined as a fraction of the swept gas mass over the initial entire torus mass. 

We solved the Riemann problem above for a range of cylindrical radii $R$, \num{32768} cells from $R = \SI{4.6}{pc}$ to $R = \SI{50}{pc}$, and then integrate the gas mass over the radii where at least one of the conditions eq.(\ref{eq:torus.velocity})--(\ref{eq:momentum.transfer}) is satisfied. 

Figure~\ref{fig:ResultsLinear}a shows the results of the analytic estimation for $f_\mathrm{strip}$. 
The stripping due to momentum transfer (the black solid curve) always predicts the highest stripped fraction. 
Models with $\eta \gtrsim 100$ predict that a significant fraction of the torus gas will be stripped away. 

Extended Data Figure~\ref{fig:ColumnDensityRatio} compares the ratio of the column-density of the infalling satellite gas $\Sigma_{\rm satellite}$ to that of the torus gas $\Sigma_{\rm torus}$ for $\eta = 100$ (solid curve) and $10$ (dotted curve). 
In models with $\eta = 100$, the column-density of the infalling gas is higher than that of the torus gas for all radii. 
On the other hand, for models with $\eta = 10$, the column-density of the infalling gas is lower compared to that of the torus gas in most regions. 
The ratio of the column-densities determines the critical value of $\eta$, underlining the importance of momentum transfer for gas stripping.

\subsection{Three-dimensional hydrodynamic simulations.}
We have developed a flat MPI parallelized code adopting the HLLC scheme\cite{Toro1994, Batten1997} for the approximate Riemann solver with the second-order MUSCL interpolation. 
We performed \num{12} runs with a grid of $256^3$ cells for $(\SI{200}{pc})^3$ box as a parameter study, a run with a grid of $1024^3$ cells for $(\SI{200}{pc})^3$ box as convergence checks and a run with a torus twice the size of the fiducial model using a grid with $512^3$ cells for $(\SI{400}{pc})^3$ box. 
Extended Data Figure~\ref{tab:parameters} summarises the parameters in the simulations. 

The computational box is aligned with the infall direction of the satellite. 
The infalling gas is modelled by a steady and uniform inflow boundary condition (the gas density is $4.5\times10^{-23} f_\mathrm{gas}$~\si{g.cm^{-3}}, the velocity is $\SI{850}{km.s^{-1}}$ and the temperature is \SI{e+4}{K}) for \SI{1.1}{Myr} from $t=0$, which is switched to a force-free boundary condition at $t = \SI{1.1}{Myr}$. 
The remaining five boundaries are force-free boundaries at all times. 
Although there is no observational correlation between the rotation axis of the torus and that of the galactic disc\cite{Schmitt2001}, we here assume that the galactic disc (with the rotation axis inclined by \SI{44.1}{deg} to the direction of the satellite infall) and the torus share a rotation axis. 
We estimate the stripped fraction of the torus $f_\mathrm{strip}$ based on the bound gas mass at $t = \SI{1.3}{Myr}$. 
Since the bound fraction $f_\mathrm{bound}$ is defined as the gas mass bound by the gravitational potential of the MBH and bulge as a fraction of the initial gas mass, $f_\mathrm{strip}$ is simply $1 - f_\mathrm{bound}$.

\subsection{Dynamical evolution of the infalling satellite in the vicinity of the galactic centre.}
Here we discuss possible processes changing the column-density of the infalling gas such as ram-pressure stripping and tidal stripping/compression. 
First of all, systematic studies concluded that the efficiency of the ram-pressure stripping within a certain period of time depends upon a total mass of the satellite galaxies, and that is, less-massive dwarfs lose their gasses quickly rather than massive dwarfs. 
A critical mass $M_\mathrm{c}$, which is the minimum mass for bearing the gas removal and keeping the most of the gas in infalling galaxies for a few giga-years, is estimated between \SI{e+9}{M_\odot} to \SI{e+10}{M_\odot} for an ambient gas density of \SI{e-4}{cm^{-3}} and a velocity of \SI{1000}{km.s^{-1}} \citep{MoriBurkert2000, Fillingham2016}. 
The mass of the GSS's progenitor lies in this marginal range. 
However, previous studies of the GSS formation indicate that this combination of the density and the high-velocity is only realised at the central high-density region in the M31, and the progenitor of the GSS takes only \SI{10}{Myr} to pass through the central \SI{4}{kpc} region\cite{Miki2014}. 
It is clear that this transit-time is too short to strip the gas from the infalling galaxies even at the environment of the galactic centre. 
Therefore, the effect of the ram-pressure stripping is negligible, and most of its gas will certainly reach the galactic centre. 
Secondly, satellite galaxies having smaller pericentres will be more heavily influenced by tidal-stretching in the direction of the galactic centre and tidal-compressing along the other two axes when passing their pericentres. 
Then such satellite galaxies could form narrower stellar streams. 
To show the effect of these processes, we newly estimated the tidal-stretching radius of infalling satellite galaxies in the Milky Way, and we got about \SI{10}{kpc} for total satellite masses of $10^7$, $10^8$ and \SI{e+9}{M_\odot} (see the vertical lines in Fig.~\ref{fig:GaiaDR2}). 
However, the tidal stripping will not significantly change the column-density of the infalling gas while it extends the duration of the gas inflow. 
On the other hand, the tidal compression reduces the cross-section of the infalling satellite, it will increase the column-density of the infalling gas and weaken the effects of the ram-pressure stripping.

\subsection{Availability of torus modelling.}
\label{sec:validation}
The half-light radius or Full-Width-at-Half-Maximum of the Mid-infrared continuum emission from the AGN torus is a measure of the size of the inner part under direct illumination from the central nucleus, and is around one to several \si{pc} \citep{Burtscher2013, Kishimoto2011, Packham2005}. 
Emission from cooler material or fluorescence lines powered by X-rays are more extended with radii of order \SI{10}{pc} \citep{Hagiwara2013, Marinucci2013}. 
For instance, the radius of the central molecular torus in the Circinus galaxy is estimated using ALMA to be about \SI{30}{pc} \citep{Izumi2018}. 
In radiative transfer calculations for torii, the outer radii are assumed to be \SIrange[range-phrase = --, range-units = single]{12}{56}{pc} \citep{Hoenig2006, Nenkova2008}. 
In our calculations, we generally adopt an outer radius of \SI{50}{pc}, and for the simulation with the larger torus, an outer radius of \SI{100}{pc}. 

Although the torus is likely composed of numerous clumps\cite{KrolikBegelman1988}, we assume a smooth distribution of gas in the torus in this study, for simplicity. 
We expect that clumps are destroyed almost instantaneously during the passage of the infalling gas and that the approximation of a smooth density distribution is appropriate. 

In order to show that each clump is destroyed immediately by the infalling satellite gas, we performed the following three-dimensional hydrodynamic simulations. 
At the outer edge of a torus, clumps have low temperatures ($\sim \SI{100}{K}$) \citep{Radovich1999}. 
At \SIrange[range-phrase = --, range-units = single]{10}{50}{pc} from the central $\SI{1.4e+8}{M_\odot}$ black hole, such a clump would have a density of $10^{-20}$--$\SI{e-17}{g.cm^{-3}}$ and a mass of several to tens $\SI{}{M_\odot}$ \citep{KawaguchiMori2011}. 
The simulations were carried out using the HLLC scheme with MUSCL interpolation and periodic boundary conditions. 
The time evolution of a clump exposed to a fast gas flow ($512 \times 512 \times 2048$ grid points for $\SI{1.6}{pc} \times \SI{1.6}{pc} \times \SI{6.4}{pc}$ box) is shown in Extended Data Figure~\ref{fig:clump}. 
The clump initially has a uniform density of $\SI{2.3e-19}{g.cm^{-3}}$ with a temperature of $\SI{e+5.5}{K}$, and the ambient flow has a density, a temperature and a velocity of $\SI{4.5e-24}{g.cm^{-3}}$, $\SI{e+4}{K}$ and $\SI{850}{km.s^{-1}}$, respectively. 
Within a timescale much shorter than that of the torus destruction (Fig.~\ref{fig:time.evolution}), a clump is deformed and expands to much lower densities. 
For a variety of clump parameters in $256 \times 256 \times 1024$ grid points simulations, we find that the time evolution of gas stripping from a clump is extensively similar (Extended Data Figure~\ref{fig:massloss}). 
In about 10 times the flow crossing timescale, a clump loses about half of its initial mass (Extended Data Figure~\ref{fig:massloss-scaled}).

\subsection{Estimation of the torus mass.}
We estimate the torus mass following Nenkova et al.\cite{Nenkova2008} under the assumption of a sharp-edge angular distribution of clumps; the torus mass is expressed as 
\begin{align}
    M_\mathrm{torus} &= 4 \pi m_\mathrm{H} N_0 N_\mathrm{H}^{(1)} \sin{\sigma} R_\mathrm{d}^2 I_q\pqty{Y},
    \label{eq:torus.mass}
    \\&
    I_q\pqty{Y} =
    \begin{cases}
        \frac{Y^2 - 1}{2 \ln{Y}} & \qfor q = 1,\\
        \frac{2 Y^2 \ln{Y}}{Y^2 - 1} & \qfor q = 3,\\
        \frac{q - 1}{q - 3} \frac{Y^{3 - q} - 1}{Y^{1 - q} - 1} & \qfor \mathrm{others},
    \end{cases}
\end{align}
where $m_\mathrm{H}$ is the hydrogen mass, $N_0$ is the average number of clumps along a radial equatorial ray, $\sigma$ is the angular width parameter, $R_\mathrm{d}$ is the inner radius, $q$ is the power-law index of the radial distribution of clumps ($r^{-q}$) and $Y \equiv R_\mathrm{out} / R_\mathrm{d}$. 
The column-density of a single clump $N_\mathrm{H}^{(1)}$ is expressed as $\SI{5.8e+21}{cm^{-2}} \times E\pqty{B - V}$, where $E\pqty{B - V} = A_V / 3.09$ and $A_V = 1.086 \tau_V$, where $\tau_V$ is the optical depth of individual clumps\cite{Bohlin1978,Rieke1985}. 

Data points plotted in Figure~\ref{fig:Mtorus-Mhost} are a compilation of observed QSOs\cite{Mor2009} and Seyfert galaxies\cite{Ichikawa2015,Fuller2016,Garcia-Bernete2019}: torus mass estimated using eq.(\ref{eq:torus.mass}) for AGN\cite{Mor2009,Ichikawa2015,Fuller2016} or torus mass of Seyfert galaxies estimated by Garcia-Bernete et al.\cite{Garcia-Bernete2019}

\subsection{Estimation of the stellar mass of AGN host galaxies.}
We estimate the stellar mass of host galaxies through their absolute magnitude in $H$ band (QSOs\cite{Veilleux2009} and Seyfert galaxies\cite{Skrutskie2006}). 
The top axis of Figure~\ref{fig:Mtorus-Mhost} indicates the corresponding stellar mass under the $M_H$--$M_*$ relation\cite{Rettura2006}.

\subsection{Estimation of the gas column-density of the infalling satellite.}
Observations show that the radial profiles of the neutral gas (atomic hydrogen, molecular hydrogen and helium) in nearby disc galaxies are universal\cite{BigielBlitz2012,Wang2014}: 
\begin{align}
    \Sigma_\mathrm{gas}(R) = 2.1 \Sigma_\mathrm{trans} \exp(-1.65 \frac{R}{R_{25}}),
\end{align}
where $\Sigma_\mathrm{trans} = \SI{14}{M_\odot.pc^{-2}}$ is the gas surface-density at the transition radius\cite{Leroy2008} (where H$_\mathrm{I}$ and H$_2$ column-densities are equal), and $R_{25}$ is the radius of the $\SI{25}{mag.arcsec^{-2}}$ $B$-band isophote. 
Under the assumption of a constant density profile in the vertical direction with thickness $z_\mathrm{d}$, 
\begin{align}
    \rho(R, z) = \frac{\Sigma_\mathrm{gas}(R)}{2 z_\mathrm{d}}
    \qfor \abs{z} \leq z_\mathrm{d}
    ,
\end{align}
the column-density along a line passing $R = 0$ and $z = 0$, inclined from the rotation axis of the disc by $\theta$, is 
\begin{align}
    \Sigma(\theta) = \frac{2.1 \Sigma_\mathrm{trans}}{1.65 \sin{\theta}} \frac{R_{25}}{z_\mathrm{d}} \bqty{1 - \exp(-1.65 \frac{z_\mathrm{d}}{R_{25}} \tan{\theta})}
    .
\end{align}
The key parameter to determine the column-density is the aspect ratio $z_\mathrm{d} / R_{25}$ and we assume $z_\mathrm{d} / R_{25} = 1 / 100$ for simplicity.

\subsection{Orbital parameters of the satellite galaxies in the Milky Way.}
\label{sec:MW.satellites}
We performed a series of test-particle simulations in a fixed potential-field of the Milky Way to estimate the pericentric distance and the orbital period of Milky Way satellites. 
The location and velocity of the satellites which include \num{12} classical dwarfs and \num{28} ultra-faint dwarfs are taken from recent observations\cite{Gaia2018,Fritz2018,McConnachie2012,Torrealba2018,Torrealba2016,Longeard2018,Koposov2018} and the NASA/IPAC Extragalactic Database (NED). 
The number of test-particles per satellite was $10^3$. 
We assumed that Sgr A* was the fixed Galactic centre by neglecting the gravitational Brownian motion of Sgr A*. 
The location of Sgr A* is $\pqty{l, b} = \SIlist[list-units = brackets, list-pair-separator={,}]{-0.056; -0.046}{degrees}$ \citep{Bland-HawthornGerhard2016} and the distance to Sgr A*, $R_0$, is $8.2467 \pm 0.009 \pm 0.045$~\si{kpc} \citep{Gravity2020}. 
The proper motion of Sgr A* is $\pqty{v_l, v_b} = \SIlist[list-units = brackets, list-pair-separator={,}, separate-uncertainty = true]{-6.369 +- 0.026; -0.202 +- 0.019}{mas.yr^{-1}}$ \citep{ReidBrunthaler2004} and the line-of-sight velocity of Sgr A* is $\SI[separate-uncertainty = true]{-7.2 +- 8.5}{km.s^{-1}}$ \citep{Reid2009}. 
The potential-field of the Milky Way model\cite{Cautun2020} (composed of an NFW halo\cite{Navarro1995,Navarro1996}, a McMillan bulge\cite{McMillan2017}, two exponential discs representing the thin and thick stellar discs, two exponential discs with a central hole\cite{KalberlaDedes2008} representing the H$_\mathrm{I}$ and H$_2$ discs and the circumgalactic medium described in a single-power-law model) is generated by an updated version of the \texttt{MAGI}\cite{MikiUmemura2018} code. 

Orbit integration backwards in time over the cosmic age was performed using the fourth-order Hermite scheme with adaptive time steps to resolve highly-eccentric orbits. 
Figure~\ref{fig:GaiaDR2} and Extended Data Figure~\ref{tab:MWsat} summarise the results of the test-particle simulations (median values with \num{84.1} and \num{15.9} percentile for the $1 \sigma$ confidence level). 
Since the estimation of the orbital parameters of test-particles unbound or having longer orbital-period than the integration time (\SI{13.8}{Gyr}) is impossible, a subset of satellites suffers from poor statistics: Leo I, Phoenix dwarf, Eridanus II, Leo IV, Leo V, Hydra II, Pisces II and Grus I. 
We exclude these satellites with an unreliable estimation of the orbital parameters from the figure and table.

\subsection{Estimation of the tidal-stripping radius of satellite galaxies.}
We estimate the tidal-stripping radius of satellite galaxies due to tidal stripping by solving the force balance between the self-gravity of the satellite and the tearing force by the central galaxy (half of the tidal force). 
The mass model of the central galaxy is the spherical-averaged Milky Way model $M_\mathrm{MW}(r)$ exploited in the previous section. 
We assume an NFW sphere as a satellite galaxy and set the concentration parameter using the $c-M$ relation\cite{IshiyamaAndo2020} as a function of the satellite mass $M_\mathrm{sat}$. 
Assuming 95\% of the initial mass of the satellite is tidally stripped, the tidal-stripping radius $r_\mathrm{tidal}$ must satisfy the following equation,
\begin{align}
    \frac{1}{2} \bqty{\frac{G M_\mathrm{MW}\pqty{r_\mathrm{tidal} - r_{5\%}}}{\pqty{r_\mathrm{tidal} - r_{5\%}}^2} - \frac{G M_\mathrm{MW}\pqty{r_\mathrm{tidal} + r_{5\%}}}{\pqty{r_\mathrm{tidal} + r_{5\%}}^2}}
    = 0.05 \frac{G M_\mathrm{sat}}{r_{5\%}^2}
    ,
\end{align}
where $r_{5\%}$ is the radius within which 5\% of the satellite mass is contained. 
Solving this equation numerically, the resultant radius $r_\mathrm{tidal}$ is \SIlist{8.41;10.4;13.0}{kpc} for satellite mass of \SIlist{e+7;e+8;e+9}{M_\odot}, respectively. 
Vertical lines in Figure~\ref{fig:GaiaDR2} indicate the tidal-stripping radius, and they locate near the lower bound for the pericentres of the current orbiting satellites.

\end{methods}

\section*{Data availability}
The data that support the findings of this study are available from the corresponding author upon reasonable request, although the requester will be responsible for providing the very considerable resources needed for transferring and storing these data.

\section*{Code availability}
The parent code \texttt{MAGI} has been publicly available at \url{https://bitbucket.org/ymiki/magi}. 
It is expected that most of the extensions and modifications made to meet the specific requirements for this project will be made available in the future release; those interested can contact the corresponding author for further information.

%% References in Methods

%%
%% EXTENDED DATA FIGURES in Methods
%%
\renewcommand{\figurename}{Extended Data Figure}

\begin{thebibliography}{10}
\expandafter\ifx\csname url\endcsname\relax
  \def\url#1{\texttt{#1}}\fi
\expandafter\ifx\csname urlprefix\endcsname\relax\def\urlprefix{URL }\fi
\providecommand{\bibinfo}[2]{#2}
\providecommand{\eprint}[2][]{\url{#2}}

\bibitem{Sanders1988}
\bibinfo{author}{{Sanders}, D.~B.} \emph{et~al.}
\newblock \bibinfo{title}{{Ultraluminous infrared galaxies and the origin of
  quasars}}.
\newblock \emph{\bibinfo{journal}{\apj}} \textbf{\bibinfo{volume}{325}},
  \bibinfo{pages}{74--91} (\bibinfo{year}{1988}).

\bibitem{HernquistMihos1995}
\bibinfo{author}{{Hernquist}, L.} \& \bibinfo{author}{{Mihos}, J.~C.}
\newblock \bibinfo{title}{{Excitation of Activity in Galaxies by Minor
  Mergers}}.
\newblock \emph{\bibinfo{journal}{\apj}} \textbf{\bibinfo{volume}{448}},
  \bibinfo{pages}{41} (\bibinfo{year}{1995}).

\bibitem{Cisternas2011}
\bibinfo{author}{{Cisternas}, M.} \emph{et~al.}
\newblock \bibinfo{title}{{The Bulk of the Black Hole Growth Since z
  \raisebox{-0.5ex}\textasciitilde 1 Occurs in a Secular Universe: No Major
  Merger-AGN Connection}}.
\newblock \emph{\bibinfo{journal}{\apj}} \textbf{\bibinfo{volume}{726}},
  \bibinfo{pages}{57} (\bibinfo{year}{2011}).

\bibitem{Miki2014}
\bibinfo{author}{{Miki}, Y.}, \bibinfo{author}{{Mori}, M.},
  \bibinfo{author}{{Kawaguchi}, T.} \& \bibinfo{author}{{Saito}, Y.}
\newblock \bibinfo{title}{{Hunting a Wandering Supermassive Black Hole in the
  M31 Halo Hermitage}}.
\newblock \emph{\bibinfo{journal}{\apj}} \textbf{\bibinfo{volume}{783}},
  \bibinfo{pages}{87} (\bibinfo{year}{2014}).

\bibitem{Miki2016}
\bibinfo{author}{{Miki}, Y.}, \bibinfo{author}{{Mori}, M.} \&
  \bibinfo{author}{{Rich}, R.~M.}
\newblock \bibinfo{title}{{Collision Tomography: Physical Properties of
  Possible Progenitors of the Andromeda Stellar Stream}}.
\newblock \emph{\bibinfo{journal}{\apj}} \textbf{\bibinfo{volume}{827}},
  \bibinfo{pages}{82} (\bibinfo{year}{2016}).

\bibitem{Lintott2009}
\bibinfo{author}{{Lintott}, C.~J.} \emph{et~al.}
\newblock \bibinfo{title}{{Galaxy Zoo: `Hanny's Voorwerp', a quasar light
  echo?}}
\newblock \emph{\bibinfo{journal}{\mnras}} \textbf{\bibinfo{volume}{399}},
  \bibinfo{pages}{129--140} (\bibinfo{year}{2009}).

\bibitem{Keel2012}
\bibinfo{author}{{Keel}, W.~C.} \emph{et~al.}
\newblock \bibinfo{title}{{The Galaxy Zoo survey for giant AGN-ionized clouds:
  past and present black hole accretion events}}.
\newblock \emph{\bibinfo{journal}{\mnras}} \textbf{\bibinfo{volume}{420}},
  \bibinfo{pages}{878--900} (\bibinfo{year}{2012}).

\bibitem{Keel2017}
\bibinfo{author}{{Keel}, W.~C.} \emph{et~al.}
\newblock \bibinfo{title}{{Fading AGN Candidates: AGN Histories and Outflow
  Signatures}}.
\newblock \emph{\bibinfo{journal}{\apj}} \textbf{\bibinfo{volume}{835}},
  \bibinfo{pages}{256} (\bibinfo{year}{2017}).

\bibitem{VillarMartin2018}
\bibinfo{author}{{Villar-Mart{\'\i}n}, M.} \emph{et~al.}
\newblock \bibinfo{title}{{A 100 kpc nebula associated with the `Teacup' fading
  quasar}}.
\newblock \emph{\bibinfo{journal}{\mnras}} \textbf{\bibinfo{volume}{474}},
  \bibinfo{pages}{2302--2312} (\bibinfo{year}{2018}).

\bibitem{Yoshida2002}
\bibinfo{author}{{Yoshida}, M.} \emph{et~al.}
\newblock \bibinfo{title}{{Discovery of a Very Extended Emission-Line Region
  around the Seyfert 2 Galaxy NGC 4388}}.
\newblock \emph{\bibinfo{journal}{\apj}} \textbf{\bibinfo{volume}{567}},
  \bibinfo{pages}{118--129} (\bibinfo{year}{2002}).

\bibitem{Kawaguchi2014}
\bibinfo{author}{{Kawaguchi}, T.}, \bibinfo{author}{{Saito}, Y.},
  \bibinfo{author}{{Miki}, Y.} \& \bibinfo{author}{{Mori}, M.}
\newblock \bibinfo{title}{{Relics of Galaxy Merging: Observational Predictions
  for a Wandering Massive Black Hole and Accompanying Star Cluster in the Halo
  of M31}}.
\newblock \emph{\bibinfo{journal}{\apjl}} \textbf{\bibinfo{volume}{789}},
  \bibinfo{pages}{L13} (\bibinfo{year}{2014}).

\bibitem{Ibata2001}
\bibinfo{author}{{Ibata}, R.}, \bibinfo{author}{{Irwin}, M.},
  \bibinfo{author}{{Lewis}, G.}, \bibinfo{author}{{Ferguson}, A. M.~N.} \&
  \bibinfo{author}{{Tanvir}, N.}
\newblock \bibinfo{title}{{A giant stream of metal-rich stars in the halo of
  the galaxy M31}}.
\newblock \emph{\bibinfo{journal}{\nat}} \textbf{\bibinfo{volume}{412}},
  \bibinfo{pages}{49--52} (\bibinfo{year}{2001}).

\bibitem{Fardal2007}
\bibinfo{author}{{Fardal}, M.~A.}, \bibinfo{author}{{Guhathakurta}, P.},
  \bibinfo{author}{{Babul}, A.} \& \bibinfo{author}{{McConnachie}, A.~W.}
\newblock \bibinfo{title}{{Investigating the Andromeda stream - III. A young
  shell system in M31}}.
\newblock \emph{\bibinfo{journal}{\mnras}} \textbf{\bibinfo{volume}{380}},
  \bibinfo{pages}{15--32} (\bibinfo{year}{2007}).

\bibitem{MoriRich2008}
\bibinfo{author}{{Mori}, M.} \& \bibinfo{author}{{Rich}, R.~M.}
\newblock \bibinfo{title}{{The Once and Future Andromeda Stream}}.
\newblock \emph{\bibinfo{journal}{\apjl}} \textbf{\bibinfo{volume}{674}},
  \bibinfo{pages}{L77--L80} (\bibinfo{year}{2008}).

\bibitem{Kirihara2017}
\bibinfo{author}{{Kirihara}, T.}, \bibinfo{author}{{Miki}, Y.},
  \bibinfo{author}{{Mori}, M.}, \bibinfo{author}{{Kawaguchi}, T.} \&
  \bibinfo{author}{{Rich}, R.~M.}
\newblock \bibinfo{title}{{Formation of the Andromeda giant stream: asymmetric
  structure and disc progenitor}}.
\newblock \emph{\bibinfo{journal}{\mnras}} \textbf{\bibinfo{volume}{464}},
  \bibinfo{pages}{3509--3525} (\bibinfo{year}{2017}).

\bibitem{Bender2005}
\bibinfo{author}{{Bender}, R.} \emph{et~al.}
\newblock \bibinfo{title}{{HST STIS Spectroscopy of the Triple Nucleus of M31:
  Two Nested Disks in Keplerian Rotation around a Supermassive Black Hole}}.
\newblock \emph{\bibinfo{journal}{\apj}} \textbf{\bibinfo{volume}{631}},
  \bibinfo{pages}{280--300} (\bibinfo{year}{2005}).

\bibitem{Li2009}
\bibinfo{author}{{Li}, Z.}, \bibinfo{author}{{Wang}, Q.~D.} \&
  \bibinfo{author}{{Wakker}, B.~P.}
\newblock \bibinfo{title}{{M31* and its circumnuclear environment}}.
\newblock \emph{\bibinfo{journal}{\mnras}} \textbf{\bibinfo{volume}{397}},
  \bibinfo{pages}{148--163} (\bibinfo{year}{2009}).

\bibitem{Ho2009}
\bibinfo{author}{{Ho}, L.~C.}
\newblock \bibinfo{title}{{Radiatively Inefficient Accretion in Nearby
  Galaxies}}.
\newblock \emph{\bibinfo{journal}{\apj}} \textbf{\bibinfo{volume}{699}},
  \bibinfo{pages}{626--637} (\bibinfo{year}{2009}).

\bibitem{Imanishi2010}
\bibinfo{author}{{Imanishi}, M.}, \bibinfo{author}{{Maiolino}, R.} \&
  \bibinfo{author}{{Nakagawa}, T.}
\newblock \bibinfo{title}{{Spitzer Infrared Low-Resolution Spectroscopic Study
  of Buried Active Galactic Nuclei in a Complete Sample of Nearby Ultraluminous
  Infrared Galaxies}}.
\newblock \emph{\bibinfo{journal}{\apj}} \textbf{\bibinfo{volume}{709}},
  \bibinfo{pages}{801--815} (\bibinfo{year}{2010}).

\bibitem{Saripalli2003}
\bibinfo{author}{{Saripalli}, L.}, \bibinfo{author}{{Subrahmanyan}, R.} \&
  \bibinfo{author}{{Udaya Shankar}, N.}
\newblock \bibinfo{title}{{Renewed Activity in the Radio Galaxy PKS B1545-321:
  Twin Edge-brightened Beams within Diffuse Radio Lobes}}.
\newblock \emph{\bibinfo{journal}{\apj}} \textbf{\bibinfo{volume}{590}},
  \bibinfo{pages}{181--191} (\bibinfo{year}{2003}).

\bibitem{Yamada2009}
\bibinfo{author}{{Yamada}, T.} \emph{et~al.}
\newblock \bibinfo{title}{{Moircs Deep Survey. III. Active Galactic Nuclei in
  Massive Galaxies at z = 2-4}}.
\newblock \emph{\bibinfo{journal}{\apj}} \textbf{\bibinfo{volume}{699}},
  \bibinfo{pages}{1354--1364} (\bibinfo{year}{2009}).

\bibitem{Benson2005}
\bibinfo{author}{{Benson}, A.~J.}
\newblock \bibinfo{title}{{Orbital parameters of infalling dark matter
  substructures}}.
\newblock \emph{\bibinfo{journal}{\mnras}} \textbf{\bibinfo{volume}{358}},
  \bibinfo{pages}{551--562} (\bibinfo{year}{2005}).

\bibitem{Khochfar2006}
\bibinfo{author}{{Khochfar}, S.} \& \bibinfo{author}{{Burkert}, A.}
\newblock \bibinfo{title}{{Orbital parameters of merging dark matter halos}}.
\newblock \emph{\bibinfo{journal}{\aap}} \textbf{\bibinfo{volume}{445}},
  \bibinfo{pages}{403--412} (\bibinfo{year}{2006}).

\bibitem{Barber2014}
\bibinfo{author}{{Barber}, C.}, \bibinfo{author}{{Starkenburg}, E.},
  \bibinfo{author}{{Navarro}, J.~F.}, \bibinfo{author}{{McConnachie}, A.~W.} \&
  \bibinfo{author}{{Fattahi}, A.}
\newblock \bibinfo{title}{{The orbital ellipticity of satellite galaxies and
  the mass of the Milky Way}}.
\newblock \emph{\bibinfo{journal}{\mnras}} \textbf{\bibinfo{volume}{437}},
  \bibinfo{pages}{959--967} (\bibinfo{year}{2014}).

\bibitem{Garrison-Kimmel2017}
\bibinfo{author}{{Garrison-Kimmel}, S.} \emph{et~al.}
\newblock \bibinfo{title}{{Not so lumpy after all: modelling the depletion of
  dark matter subhaloes by Milky Way-like galaxies}}.
\newblock \emph{\bibinfo{journal}{\mnras}} \textbf{\bibinfo{volume}{471}},
  \bibinfo{pages}{1709--1727} (\bibinfo{year}{2017}).

\bibitem{Morinaga2019}
\bibinfo{author}{{Morinaga}, Y.}, \bibinfo{author}{{Ishiyama}, T.},
  \bibinfo{author}{{Kirihara}, T.} \& \bibinfo{author}{{Kinjo}, K.}
\newblock \bibinfo{title}{{Statistical properties of substructures around Milky
  Way-sized haloes and their implications for the formation of stellar
  streams}}.
\newblock \emph{\bibinfo{journal}{\mnras}} \textbf{\bibinfo{volume}{487}},
  \bibinfo{pages}{2718--2729} (\bibinfo{year}{2019}).

\bibitem{Helmi2018}
\bibinfo{author}{{Helmi}, A.} \emph{et~al.}
\newblock \bibinfo{title}{{The merger that led to the formation of the Milky
  Way's inner stellar halo and thick disk}}.
\newblock \emph{\bibinfo{journal}{\nat}} \textbf{\bibinfo{volume}{563}},
  \bibinfo{pages}{85--88} (\bibinfo{year}{2018}).

\bibitem{Belokurov2018}
\bibinfo{author}{{Belokurov}, V.}, \bibinfo{author}{{Erkal}, D.},
  \bibinfo{author}{{Evans}, N.~W.}, \bibinfo{author}{{Koposov}, S.~E.} \&
  \bibinfo{author}{{Deason}, A.~J.}
\newblock \bibinfo{title}{{Co-formation of the disc and the stellar halo}}.
\newblock \emph{\bibinfo{journal}{\mnras}} \textbf{\bibinfo{volume}{478}},
  \bibinfo{pages}{611--619} (\bibinfo{year}{2018}).

\bibitem{Gordon2006}
\bibinfo{author}{{Gordon}, K.~D.} \emph{et~al.}
\newblock \bibinfo{title}{{Spitzer MIPS Infrared Imaging of M31: Further
  Evidence for a Spiral-Ring Composite Structure}}.
\newblock \emph{\bibinfo{journal}{\apjl}} \textbf{\bibinfo{volume}{638}},
  \bibinfo{pages}{L87--L92} (\bibinfo{year}{2006}).

\bibitem{RamonFox2020}
\bibinfo{author}{{Ram{\'o}n-Fox}, F.~G.} \& \bibinfo{author}{{Aceves}, H.}
\newblock \bibinfo{title}{{Accretion of small satellites and gas inflows in a
  disc galaxy}}.
\newblock \emph{\bibinfo{journal}{\mnras}} \textbf{\bibinfo{volume}{491}},
  \bibinfo{pages}{3908--3922} (\bibinfo{year}{2020}).

\end{thebibliography}

\begin{thebibliography}{10}
\expandafter\ifx\csname url\endcsname\relax
  \def\url#1{\texttt{#1}}\fi
\expandafter\ifx\csname urlprefix\endcsname\relax\def\urlprefix{URL }\fi
\providecommand{\bibinfo}[2]{#2}
\providecommand{\eprint}[2][]{\url{#2}}

\setcounter{enumiv}{30}

\bibitem{BeckertDuschl2004}
\bibinfo{author}{{Beckert}, T.} \& \bibinfo{author}{{Duschl}, W.~J.}
\newblock \bibinfo{title}{{The dynamical state of a thick cloudy torus around
  an AGN}}.
\newblock \emph{\bibinfo{journal}{\aap}} \textbf{\bibinfo{volume}{426}},
  \bibinfo{pages}{445--454} (\bibinfo{year}{2004}).

\bibitem{DiMatteo2005}
\bibinfo{author}{{Di Matteo}, T.}, \bibinfo{author}{{Springel}, V.} \&
  \bibinfo{author}{{Hernquist}, L.}
\newblock \bibinfo{title}{{Energy input from quasars regulates the growth and
  activity of black holes and their host galaxies}}.
\newblock \emph{\bibinfo{journal}{\nat}} \textbf{\bibinfo{volume}{433}},
  \bibinfo{pages}{604--607} (\bibinfo{year}{2005}).

\bibitem{GaborBournaud2014}
\bibinfo{author}{{Gabor}, J.~M.} \& \bibinfo{author}{{Bournaud}, F.}
\newblock \bibinfo{title}{{Active galactic nuclei-driven outflows without
  immediate quenching in simulations of high-redshift disc galaxies}}.
\newblock \emph{\bibinfo{journal}{\mnras}} \textbf{\bibinfo{volume}{441}},
  \bibinfo{pages}{1615--1627} (\bibinfo{year}{2014}).

\bibitem{Athanassoula2005}
\bibinfo{author}{{Athanassoula}, E.}, \bibinfo{author}{{Lambert}, J.~C.} \&
  \bibinfo{author}{{Dehnen}, W.}
\newblock \bibinfo{title}{{Can bars be destroyed by a central mass
  concentration?— I. Simulations}}.
\newblock \emph{\bibinfo{journal}{Monthly Notices of the Royal Astronomical
  Society}} \textbf{\bibinfo{volume}{363}}, \bibinfo{pages}{496--508}
  (\bibinfo{year}{2005}).

\bibitem{HopkinsQuataert2010}
\bibinfo{author}{{Hopkins}, P.~F.} \& \bibinfo{author}{{Quataert}, E.}
\newblock \bibinfo{title}{{How do massive black holes get their gas?}}
\newblock \emph{\bibinfo{journal}{\mnras}} \textbf{\bibinfo{volume}{407}},
  \bibinfo{pages}{1529--1564} (\bibinfo{year}{2010}).

\bibitem{Angles-Alcazar2020}
\bibinfo{author}{{Angles-Alcazar}, D.} \emph{et~al.}
\newblock \bibinfo{title}{{Cosmological simulations of quasar fueling to
  sub-parsec scales using Lagrangian hyper-refinement}}.
\newblock \emph{\bibinfo{journal}{arXiv e-prints}}
  \bibinfo{pages}{arXiv:2008.12303} (\bibinfo{year}{2020}).

\bibitem{Geehan2006}
\bibinfo{author}{{Geehan}, J.~J.}, \bibinfo{author}{{Fardal}, M.~A.},
  \bibinfo{author}{{Babul}, A.} \& \bibinfo{author}{{Guhathakurta}, P.}
\newblock \bibinfo{title}{{Investigating the Andromeda stream - I. Simple
  analytic bulge-disc-halo model for M31}}.
\newblock \emph{\bibinfo{journal}{\mnras}} \textbf{\bibinfo{volume}{366}},
  \bibinfo{pages}{996--1011} (\bibinfo{year}{2006}).

\bibitem{TomisakaIkeuchi1988}
\bibinfo{author}{{Tomisaka}, K.} \& \bibinfo{author}{{Ikeuchi}, S.}
\newblock \bibinfo{title}{{Starburst nucleus - Galactic-scale bipolar flow}}.
\newblock \emph{\bibinfo{journal}{\apj}} \textbf{\bibinfo{volume}{330}},
  \bibinfo{pages}{695--717} (\bibinfo{year}{1988}).

\bibitem{Okada1989}
\bibinfo{author}{{Okada}, R.}, \bibinfo{author}{{Fukue}, J.} \&
  \bibinfo{author}{{Matsumoto}, R.}
\newblock \bibinfo{title}{{A model of astrophysical tori with magnetic
  fields}}.
\newblock \emph{\bibinfo{journal}{\pasj}} \textbf{\bibinfo{volume}{41}},
  \bibinfo{pages}{133--140} (\bibinfo{year}{1989}).

\bibitem{FukueSanbuichi1993}
\bibinfo{author}{{Fukue}, J.} \& \bibinfo{author}{{Sanbuichi}, K.}
\newblock \bibinfo{title}{{Model of obscuring molecular tori in Seyfert
  nuclei}}.
\newblock \emph{\bibinfo{journal}{\pasj}} \textbf{\bibinfo{volume}{45}},
  \bibinfo{pages}{135--138} (\bibinfo{year}{1993}).

\bibitem{Bulbul2010}
\bibinfo{author}{{Bulbul}, G.~E.}, \bibinfo{author}{{Hasler}, N.},
  \bibinfo{author}{{Bonamente}, M.} \& \bibinfo{author}{{Joy}, M.}
\newblock \bibinfo{title}{{An Analytic Model of the Physical Properties of
  Galaxy Clusters}}.
\newblock \emph{\bibinfo{journal}{\apj}} \textbf{\bibinfo{volume}{720}},
  \bibinfo{pages}{1038--1044} (\bibinfo{year}{2010}).

\bibitem{Hoenig2006}
\bibinfo{author}{{H{\"o}nig}, S.~F.}, \bibinfo{author}{{Beckert}, T.},
  \bibinfo{author}{{Ohnaka}, K.} \& \bibinfo{author}{{Weigelt}, G.}
\newblock \bibinfo{title}{{Radiative transfer modeling of three-dimensional
  clumpy AGN tori and its application to NGC 1068}}.
\newblock \emph{\bibinfo{journal}{\aap}} \textbf{\bibinfo{volume}{452}},
  \bibinfo{pages}{459--471} (\bibinfo{year}{2006}).

\bibitem{KrolikBegelman1988}
\bibinfo{author}{{Krolik}, J.~H.} \& \bibinfo{author}{{Begelman}, M.~C.}
\newblock \bibinfo{title}{{Molecular tori in Seyfert galaxies - Feeding the
  monster and hiding it}}.
\newblock \emph{\bibinfo{journal}{\apj}} \textbf{\bibinfo{volume}{329}},
  \bibinfo{pages}{702--711} (\bibinfo{year}{1988}).

\bibitem{Mor2009}
\bibinfo{author}{{Mor}, R.}, \bibinfo{author}{{Netzer}, H.} \&
  \bibinfo{author}{{Elitzur}, M.}
\newblock \bibinfo{title}{{Dusty Structure Around Type-I Active Galactic
  Nuclei: Clumpy Torus Narrow-line Region and Near-nucleus Hot Dust}}.
\newblock \emph{\bibinfo{journal}{\apj}} \textbf{\bibinfo{volume}{705}},
  \bibinfo{pages}{298--313} (\bibinfo{year}{2009}).

\bibitem{Schartmann2005}
\bibinfo{author}{{Schartmann}, M.}, \bibinfo{author}{{Meisenheimer}, K.},
  \bibinfo{author}{{Camenzind}, M.}, \bibinfo{author}{{Wolf}, S.} \&
  \bibinfo{author}{{Henning}, T.}
\newblock \bibinfo{title}{{Towards a physical model of dust tori in Active
  Galactic Nuclei. Radiative transfer calculations for a hydrostatic torus
  model}}.
\newblock \emph{\bibinfo{journal}{\aap}} \textbf{\bibinfo{volume}{437}},
  \bibinfo{pages}{861--881} (\bibinfo{year}{2005}).

\bibitem{Mateo1998}
\bibinfo{author}{{Mateo}, M.~L.}
\newblock \bibinfo{title}{{Dwarf Galaxies of the Local Group}}.
\newblock \emph{\bibinfo{journal}{\araa}} \textbf{\bibinfo{volume}{36}},
  \bibinfo{pages}{435--506} (\bibinfo{year}{1998}).

\bibitem{Conselice2003}
\bibinfo{author}{{Conselice}, C.~J.}, \bibinfo{author}{{O'Neil}, K.},
  \bibinfo{author}{{Gallagher}, J.~S.} \& \bibinfo{author}{{Wyse}, R.~F.~G.}
\newblock \bibinfo{title}{{Galaxy Populations and Evolution in Clusters. IV.
  Deep H I Observations of Dwarf Elliptical Galaxies in the Virgo Cluster}}.
\newblock \emph{\bibinfo{journal}{\apj}} \textbf{\bibinfo{volume}{591}},
  \bibinfo{pages}{167--184} (\bibinfo{year}{2003}).

\bibitem{Thuan2016}
\bibinfo{author}{{Thuan}, T.~X.}, \bibinfo{author}{{Goehring}, K.~M.},
  \bibinfo{author}{{Hibbard}, J.~E.}, \bibinfo{author}{{Izotov}, Y.~I.} \&
  \bibinfo{author}{{Hunt}, L.~K.}
\newblock \bibinfo{title}{{The H I content of extremely metal-deficient blue
  compact dwarf galaxies}}.
\newblock \emph{\bibinfo{journal}{\mnras}} \textbf{\bibinfo{volume}{463}},
  \bibinfo{pages}{4268--4286} (\bibinfo{year}{2016}).

\bibitem{Toro1994}
\bibinfo{author}{{Toro}, E.~F.}, \bibinfo{author}{{Spruce}, M.} \&
  \bibinfo{author}{{Speares}, W.}
\newblock \bibinfo{title}{{Restoration of the contact surface in the
  HLL-Riemann solver}}.
\newblock \emph{\bibinfo{journal}{Shock Waves}} \textbf{\bibinfo{volume}{4}},
  \bibinfo{pages}{25--34} (\bibinfo{year}{1994}).

\bibitem{Batten1997}
\bibinfo{author}{{Batten}, P.}, \bibinfo{author}{{Clarke}, N.},
  \bibinfo{author}{{Lambert}, C.} \& \bibinfo{author}{{Causon}, D.}
\newblock \bibinfo{title}{On the choice of wavespeeds for the hllc riemann
  solver}.
\newblock \emph{\bibinfo{journal}{SIAM Journal on Scientific Computing}}
  \textbf{\bibinfo{volume}{18}}, \bibinfo{pages}{1553--1570}
  (\bibinfo{year}{1997}).

\bibitem{Schmitt2001}
\bibinfo{author}{{Schmitt}, H.~R.} \emph{et~al.}
\newblock \bibinfo{title}{{Testing the Unified Model with an Infrared-selected
  Sample of Seyfert Galaxies}}.
\newblock \emph{\bibinfo{journal}{\apj}} \textbf{\bibinfo{volume}{555}},
  \bibinfo{pages}{663--672} (\bibinfo{year}{2001}).

\bibitem{MoriBurkert2000}
\bibinfo{author}{{Mori}, M.} \& \bibinfo{author}{{Burkert}, A.}
\newblock \bibinfo{title}{{Gas Stripping of Dwarf Galaxies in Clusters of
  Galaxies}}.
\newblock \emph{\bibinfo{journal}{\apj}} \textbf{\bibinfo{volume}{538}},
  \bibinfo{pages}{559--568} (\bibinfo{year}{2000}).

\bibitem{Fillingham2016}
\bibinfo{author}{{Fillingham}, S.~P.} \emph{et~al.}
\newblock \bibinfo{title}{{Under pressure: quenching star formation in low-mass
  satellite galaxies via stripping}}.
\newblock \emph{\bibinfo{journal}{\mnras}} \textbf{\bibinfo{volume}{463}},
  \bibinfo{pages}{1916--1928} (\bibinfo{year}{2016}).

\bibitem{Burtscher2013}
\bibinfo{author}{{Burtscher}, L.} \emph{et~al.}
\newblock \bibinfo{title}{{A diversity of dusty AGN tori. Data release for the
  VLTI/MIDI AGN Large Program and first results for 23 galaxies}}.
\newblock \emph{\bibinfo{journal}{\aap}} \textbf{\bibinfo{volume}{558}},
  \bibinfo{pages}{A149} (\bibinfo{year}{2013}).

\bibitem{Kishimoto2011}
\bibinfo{author}{{Kishimoto}, M.} \emph{et~al.}
\newblock \bibinfo{title}{{Mapping the radial structure of AGN tori}}.
\newblock \emph{\bibinfo{journal}{\aap}} \textbf{\bibinfo{volume}{536}},
  \bibinfo{pages}{A78} (\bibinfo{year}{2011}).

\bibitem{Packham2005}
\bibinfo{author}{{Packham}, C.} \emph{et~al.}
\newblock \bibinfo{title}{{The Extended Mid-Infrared Structure of the Circinus
  Galaxy}}.
\newblock \emph{\bibinfo{journal}{\apjl}} \textbf{\bibinfo{volume}{618}},
  \bibinfo{pages}{L17--L20} (\bibinfo{year}{2005}).

\bibitem{Hagiwara2013}
\bibinfo{author}{{Hagiwara}, Y.}, \bibinfo{author}{{Miyoshi}, M.},
  \bibinfo{author}{{Doi}, A.} \& \bibinfo{author}{{Horiuchi}, S.}
\newblock \bibinfo{title}{{Submillimeter H$_{2}$O Maser in Circinus
  Galaxy{\textemdash}a New Probe for the Circumnuclear Region of Active
  Galactic Nuclei}}.
\newblock \emph{\bibinfo{journal}{\apjl}} \textbf{\bibinfo{volume}{768}},
  \bibinfo{pages}{L38} (\bibinfo{year}{2013}).

\bibitem{Marinucci2013}
\bibinfo{author}{{Marinucci}, A.}, \bibinfo{author}{{Miniutti}, G.},
  \bibinfo{author}{{Bianchi}, S.}, \bibinfo{author}{{Matt}, G.} \&
  \bibinfo{author}{{Risaliti}, G.}
\newblock \bibinfo{title}{{A Chandra view of the clumpy reflector at the heart
  of the Circinus galaxy}}.
\newblock \emph{\bibinfo{journal}{\mnras}} \textbf{\bibinfo{volume}{436}},
  \bibinfo{pages}{2500--2504} (\bibinfo{year}{2013}).

\bibitem{Izumi2018}
\bibinfo{author}{{Izumi}, T.}, \bibinfo{author}{{Wada}, K.},
  \bibinfo{author}{{Fukushige}, R.}, \bibinfo{author}{{Hamamura}, S.} \&
  \bibinfo{author}{{Kohno}, K.}
\newblock \bibinfo{title}{{Circumnuclear Multiphase Gas in the Circinus Galaxy.
  II. The Molecular and Atomic Obscuring Structures Revealed with ALMA}}.
\newblock \emph{\bibinfo{journal}{\apj}} \textbf{\bibinfo{volume}{867}},
  \bibinfo{pages}{48} (\bibinfo{year}{2018}).

\bibitem{Nenkova2008}
\bibinfo{author}{{Nenkova}, M.}, \bibinfo{author}{{Sirocky}, M.~M.},
  \bibinfo{author}{{Nikutta}, R.}, \bibinfo{author}{{Ivezi{\'c}}, {\v Z}.} \&
  \bibinfo{author}{{Elitzur}, M.}
\newblock \bibinfo{title}{{AGN Dusty Tori. II. Observational Implications of
  Clumpiness}}.
\newblock \emph{\bibinfo{journal}{\apj}} \textbf{\bibinfo{volume}{685}},
  \bibinfo{pages}{160--180} (\bibinfo{year}{2008}).

\bibitem{Radovich1999}
\bibinfo{author}{{Radovich}, M.}, \bibinfo{author}{{Klaas}, U.},
  \bibinfo{author}{{Acosta-Pulido}, J.} \& \bibinfo{author}{{Lemke}, D.}
\newblock \bibinfo{title}{{The 10-200 MU M spectral energy distribution of the
  prototype narrow-line X-ray galaxy NGC 7582}}.
\newblock \emph{\bibinfo{journal}{\aap}} \textbf{\bibinfo{volume}{348}},
  \bibinfo{pages}{705--710} (\bibinfo{year}{1999}).

\bibitem{KawaguchiMori2011}
\bibinfo{author}{{Kawaguchi}, T.} \& \bibinfo{author}{{Mori}, M.}
\newblock \bibinfo{title}{{Near-infrared Reverberation by Dusty Clumpy Tori in
  Active Galactic Nuclei}}.
\newblock \emph{\bibinfo{journal}{\apj}} \textbf{\bibinfo{volume}{737}},
  \bibinfo{pages}{105} (\bibinfo{year}{2011}).

\bibitem{Bohlin1978}
\bibinfo{author}{{Bohlin}, R.~C.}, \bibinfo{author}{{Savage}, B.~D.} \&
  \bibinfo{author}{{Drake}, J.~F.}
\newblock \bibinfo{title}{{A survey of interstellar H I from Lalpha absorption
  measurements. II.}}
\newblock \emph{\bibinfo{journal}{\apj}} \textbf{\bibinfo{volume}{224}},
  \bibinfo{pages}{132--142} (\bibinfo{year}{1978}).

\bibitem{Rieke1985}
\bibinfo{author}{{Rieke}, G.~H.} \& \bibinfo{author}{{Lebofsky}, M.~J.}
\newblock \bibinfo{title}{{The interstellar extinction law from 1 to 13
  microns.}}
\newblock \emph{\bibinfo{journal}{\apj}} \textbf{\bibinfo{volume}{288}},
  \bibinfo{pages}{618--621} (\bibinfo{year}{1985}).

\bibitem{Ichikawa2015}
\bibinfo{author}{{Ichikawa}, K.} \emph{et~al.}
\newblock \bibinfo{title}{{The Differences in the Torus Geometry between Hidden
  and Non-hidden Broad Line Active Galactic Nuclei}}.
\newblock \emph{\bibinfo{journal}{\apj}} \textbf{\bibinfo{volume}{803}},
  \bibinfo{pages}{57} (\bibinfo{year}{2015}).

\bibitem{Fuller2016}
\bibinfo{author}{{Fuller}, L.} \emph{et~al.}
\newblock \bibinfo{title}{{Investigating the dusty torus of Seyfert galaxies
  using SOFIA/FORCAST photometry}}.
\newblock \emph{\bibinfo{journal}{\mnras}} \textbf{\bibinfo{volume}{462}},
  \bibinfo{pages}{2618--2630} (\bibinfo{year}{2016}).

\bibitem{Garcia-Bernete2019}
\bibinfo{author}{{Garc{\'\i}a-Bernete}, I.} \emph{et~al.}
\newblock \bibinfo{title}{{Torus model properties of an ultra-hard X-ray
  selected sample of Seyfert galaxies}}.
\newblock \emph{\bibinfo{journal}{\mnras}} \textbf{\bibinfo{volume}{486}},
  \bibinfo{pages}{4917--4935} (\bibinfo{year}{2019}).

\bibitem{Veilleux2009}
\bibinfo{author}{{Veilleux}, S.} \emph{et~al.}
\newblock \bibinfo{title}{{A Deep Hubble Space Telescope H-Band Imaging Survey
  of Massive Gas-Rich Mergers. II. The QUEST QSOs}}.
\newblock \emph{\bibinfo{journal}{\apj}} \textbf{\bibinfo{volume}{701}},
  \bibinfo{pages}{587--606} (\bibinfo{year}{2009}).

\bibitem{Skrutskie2006}
\bibinfo{author}{{Skrutskie}, M.~F.} \emph{et~al.}
\newblock \bibinfo{title}{{The Two Micron All Sky Survey (2MASS)}}.
\newblock \emph{\bibinfo{journal}{\aj}} \textbf{\bibinfo{volume}{131}},
  \bibinfo{pages}{1163--1183} (\bibinfo{year}{2006}).

\bibitem{Rettura2006}
\bibinfo{author}{{Rettura}, A.} \emph{et~al.}
\newblock \bibinfo{title}{{Comparing dynamical and photometric-stellar masses
  of early-type galaxies at z \raisebox{-0.5ex}\textasciitilde 1}}.
\newblock \emph{\bibinfo{journal}{\aap}} \textbf{\bibinfo{volume}{458}},
  \bibinfo{pages}{717--726} (\bibinfo{year}{2006}).

\bibitem{BigielBlitz2012}
\bibinfo{author}{{Bigiel}, F.} \& \bibinfo{author}{{Blitz}, L.}
\newblock \bibinfo{title}{{A Universal Neutral Gas Profile for nearby Disk
  Galaxies}}.
\newblock \emph{\bibinfo{journal}{\apj}} \textbf{\bibinfo{volume}{756}},
  \bibinfo{pages}{183} (\bibinfo{year}{2012}).

\bibitem{Wang2014}
\bibinfo{author}{{Wang}, J.} \emph{et~al.}
\newblock \bibinfo{title}{{An observational and theoretical view of the radial
  distribution of H I gas in galaxies}}.
\newblock \emph{\bibinfo{journal}{\mnras}} \textbf{\bibinfo{volume}{441}},
  \bibinfo{pages}{2159--2172} (\bibinfo{year}{2014}).

\bibitem{Leroy2008}
\bibinfo{author}{{Leroy}, A.~K.} \emph{et~al.}
\newblock \bibinfo{title}{{The Star Formation Efficiency in Nearby Galaxies:
  Measuring Where Gas Forms Stars Effectively}}.
\newblock \emph{\bibinfo{journal}{\aj}} \textbf{\bibinfo{volume}{136}},
  \bibinfo{pages}{2782--2845} (\bibinfo{year}{2008}).

\bibitem{Gaia2018}
\bibinfo{author}{{Gaia Collaboration}} \emph{et~al.}
\newblock \bibinfo{title}{{Gaia Data Release 2. Kinematics of globular clusters
  and dwarf galaxies around the Milky Way}}.
\newblock \emph{\bibinfo{journal}{\aap}} \textbf{\bibinfo{volume}{616}},
  \bibinfo{pages}{A12} (\bibinfo{year}{2018}).

\bibitem{Fritz2018}
\bibinfo{author}{{Fritz}, T.~K.} \emph{et~al.}
\newblock \bibinfo{title}{{Gaia DR2 proper motions of dwarf galaxies within 420
  kpc. Orbits, Milky Way mass, tidal influences, planar alignments, and group
  infall}}.
\newblock \emph{\bibinfo{journal}{\aap}} \textbf{\bibinfo{volume}{619}},
  \bibinfo{pages}{A103} (\bibinfo{year}{2018}).

\bibitem{McConnachie2012}
\bibinfo{author}{{McConnachie}, A.~W.}
\newblock \bibinfo{title}{{The Observed Properties of Dwarf Galaxies in and
  around the Local Group}}.
\newblock \emph{\bibinfo{journal}{\aj}} \textbf{\bibinfo{volume}{144}},
  \bibinfo{pages}{4} (\bibinfo{year}{2012}).

\bibitem{Torrealba2018}
\bibinfo{author}{{Torrealba}, G.} \emph{et~al.}
\newblock \bibinfo{title}{{Discovery of two neighbouring satellites in the
  Carina constellation with MagLiteS}}.
\newblock \emph{\bibinfo{journal}{\mnras}} \textbf{\bibinfo{volume}{475}},
  \bibinfo{pages}{5085--5097} (\bibinfo{year}{2018}).

\bibitem{Torrealba2016}
\bibinfo{author}{{Torrealba}, G.}, \bibinfo{author}{{Koposov}, S.~E.},
  \bibinfo{author}{{Belokurov}, V.} \& \bibinfo{author}{{Irwin}, M.}
\newblock \bibinfo{title}{{The feeble giant. Discovery of a large and diffuse
  Milky Way dwarf galaxy in the constellation of Crater}}.
\newblock \emph{\bibinfo{journal}{\mnras}} \textbf{\bibinfo{volume}{459}},
  \bibinfo{pages}{2370--2378} (\bibinfo{year}{2016}).

\bibitem{Longeard2018}
\bibinfo{author}{{Longeard}, N.} \emph{et~al.}
\newblock \bibinfo{title}{{Pristine dwarf galaxy survey - I. A detailed
  photometric and spectroscopic study of the very metal-poor Draco II
  satellite}}.
\newblock \emph{\bibinfo{journal}{\mnras}} \textbf{\bibinfo{volume}{480}},
  \bibinfo{pages}{2609--2627} (\bibinfo{year}{2018}).

\bibitem{Koposov2018}
\bibinfo{author}{{Koposov}, S.~E.} \emph{et~al.}
\newblock \bibinfo{title}{{Snake in the Clouds: a new nearby dwarf galaxy in
  the Magellanic bridge*}}.
\newblock \emph{\bibinfo{journal}{\mnras}} \textbf{\bibinfo{volume}{479}},
  \bibinfo{pages}{5343--5361} (\bibinfo{year}{2018}).

\bibitem{Bland-HawthornGerhard2016}
\bibinfo{author}{{Bland-Hawthorn}, J.} \& \bibinfo{author}{{Gerhard}, O.}
\newblock \bibinfo{title}{{The Galaxy in Context: Structural, Kinematic, and
  Integrated Properties}}.
\newblock \emph{\bibinfo{journal}{\araa}} \textbf{\bibinfo{volume}{54}},
  \bibinfo{pages}{529--596} (\bibinfo{year}{2016}).

\bibitem{Gravity2020}
\bibinfo{author}{{Gravity Collaboration}} \emph{et~al.}
\newblock \bibinfo{title}{{Detection of the Schwarzschild precession in the
  orbit of the star S2 near the Galactic centre massive black hole}}.
\newblock \emph{\bibinfo{journal}{\aap}} \textbf{\bibinfo{volume}{636}},
  \bibinfo{pages}{L5} (\bibinfo{year}{2020}).

\bibitem{ReidBrunthaler2004}
\bibinfo{author}{{Reid}, M.~J.} \& \bibinfo{author}{{Brunthaler}, A.}
\newblock \bibinfo{title}{{The Proper Motion of Sagittarius A*. II. The Mass of
  Sagittarius A*}}.
\newblock \emph{\bibinfo{journal}{\apj}} \textbf{\bibinfo{volume}{616}},
  \bibinfo{pages}{872--884} (\bibinfo{year}{2004}).

\bibitem{Reid2009}
\bibinfo{author}{{Reid}, M.~J.} \emph{et~al.}
\newblock \bibinfo{title}{{Trigonometric Parallaxes of Massive Star-Forming
  Regions. VI. Galactic Structure, Fundamental Parameters, and Noncircular
  Motions}}.
\newblock \emph{\bibinfo{journal}{\apj}} \textbf{\bibinfo{volume}{700}},
  \bibinfo{pages}{137--148} (\bibinfo{year}{2009}).

\bibitem{Cautun2020}
\bibinfo{author}{{Cautun}, M.} \emph{et~al.}
\newblock \bibinfo{title}{{The milky way total mass profile as inferred from
  Gaia DR2}}.
\newblock \emph{\bibinfo{journal}{\mnras}} \textbf{\bibinfo{volume}{494}},
  \bibinfo{pages}{4291--4313} (\bibinfo{year}{2020}).

\bibitem{Navarro1995}
\bibinfo{author}{{Navarro}, J.~F.}, \bibinfo{author}{{Frenk}, C.~S.} \&
  \bibinfo{author}{{White}, S.~D.~M.}
\newblock \bibinfo{title}{{Simulations of X-ray clusters}}.
\newblock \emph{\bibinfo{journal}{\mnras}} \textbf{\bibinfo{volume}{275}},
  \bibinfo{pages}{720--740} (\bibinfo{year}{1995}).

\bibitem{Navarro1996}
\bibinfo{author}{{Navarro}, J.~F.}, \bibinfo{author}{{Frenk}, C.~S.} \&
  \bibinfo{author}{{White}, S.~D.~M.}
\newblock \bibinfo{title}{{The Structure of Cold Dark Matter Halos}}.
\newblock \emph{\bibinfo{journal}{\apj}} \textbf{\bibinfo{volume}{462}},
  \bibinfo{pages}{563} (\bibinfo{year}{1996}).

\bibitem{McMillan2017}
\bibinfo{author}{{McMillan}, P.~J.}
\newblock \bibinfo{title}{{The mass distribution and gravitational potential of
  the Milky Way}}.
\newblock \emph{\bibinfo{journal}{\mnras}} \textbf{\bibinfo{volume}{465}},
  \bibinfo{pages}{76--94} (\bibinfo{year}{2017}).

\bibitem{KalberlaDedes2008}
\bibinfo{author}{{Kalberla}, P.~M.~W.} \& \bibinfo{author}{{Dedes}, L.}
\newblock \bibinfo{title}{{Global properties of the H I distribution in the
  outer Milky Way. Planar and extra-planar gas}}.
\newblock \emph{\bibinfo{journal}{\aap}} \textbf{\bibinfo{volume}{487}},
  \bibinfo{pages}{951--963} (\bibinfo{year}{2008}).

\bibitem{MikiUmemura2018}
\bibinfo{author}{{Miki}, Y.} \& \bibinfo{author}{{Umemura}, M.}
\newblock \bibinfo{title}{{MAGI: many-component galaxy initializer}}.
\newblock \emph{\bibinfo{journal}{\mnras}} \textbf{\bibinfo{volume}{475}},
  \bibinfo{pages}{2269--2281} (\bibinfo{year}{2018}).

\bibitem{IshiyamaAndo2020}
\bibinfo{author}{{Ishiyama}, T.} \& \bibinfo{author}{{Ando}, S.}
\newblock \bibinfo{title}{{The abundance and structure of subhaloes near the
  free streaming scale and their impact on indirect dark matter searches}}.
\newblock \emph{\bibinfo{journal}{\mnras}} \textbf{\bibinfo{volume}{492}},
  \bibinfo{pages}{3662--3671} (\bibinfo{year}{2020}).

\end{thebibliography}
\end{document}